\documentclass[10pt,twocolumn]{article}

\usepackage[margin=0.75in,columnsep=0.24in]{geometry}
\usepackage[T1]{fontenc}
\usepackage{lmodern}
\usepackage{microtype}

\usepackage{graphicx}
\usepackage{amsmath}
\usepackage{amssymb}      
\usepackage{makecell}
\usepackage{url}
\usepackage{booktabs}     
\usepackage{multirow}     
\usepackage{array}        
\usepackage{threeparttable} 
\usepackage{float} 
\usepackage{subcaption} 
\usepackage{xcolor}       
\usepackage[colorlinks=true,linkcolor=blue,citecolor=blue,urlcolor=blue]{hyperref} 
\usepackage[capitalize]{cleveref} 
\usepackage{tikz}

\usetikzlibrary{shapes.geometric, arrows.meta, positioning, calc, fit, backgrounds, quotes}

\begin{document}

\title{A Time-Multiplexed Spiking Neural Network Accelerator with Pipelined Readout for FPGA Inference}

\author{Reza Ansari \quad and \quad Maciej Wielgosz\\
\small AGH University of Science and Technology, Kraków, Poland\\
\small \texttt{m.rezaansari71@gmail.com; wielgosz@agh.edu.pl}}
\date{}

\maketitle

\begin{abstract}
Spiking Neural Networks (SNNs) provide a power-efficient, neuromorphic alternative to traditional artificial neural networks by processing information via discrete temporal events. This paper presents the design and Field Programmable Gate Array (FPGA) implementation of an inference-only SNN accelerator optimized for MNIST digit classification. To overcome physical routing constraints and timing bottlenecks inherent in low-cost devices, we propose a highly optimized hardware microarchitecture. This architecture features a time-multiplexed 1-bit spike-feeding mechanism governed by a finite state machine (FSM) controller, localized distributed memory for weight storage, and an integer-friendly Leaky Integrate-and-Fire (LIF) neuron model with optimized register widths to prevent overflow hazards. Furthermore, a multi-cycle pipelined argmax and tie-breaker readout module is introduced to eliminate the dominant combinational critical path. Implemented on an entry-level AMD Artix-7 FPGA (XC7A200T) using a 784-64-10 network topology, the proposed pipelined architecture increases the maximum operating frequency ($F_{max}$) from 13.3 MHz to 167 MHz. Hardware evaluations demonstrate a sequential processing latency of 82 $\mu$s per image, completing a 1,000-sample VHDL simulation test batch in 0.082 s. Vivado post-implementation vector-based power analysis yields an estimated total on-chip power consumption of 0.336 W, achieving an estimated energy efficiency of approximately 36,300 samples per Joule.
These results highlight the proposed microarchitecture as a resource-efficient solution for real-time neuromorphic edge inference, provided the network size remains within the practical bounds of its time-multiplexed execution.
\end{abstract}


\section{Introduction}
\label{sec:introduction}
Spiking Neural Networks (SNNs) represent a significant shift toward neuromorphic hardware design, offering a biologically inspired alternative to traditional continuous-activation models. By processing information through discrete temporal events, SNNs can significantly reduce computational overhead and power consumption during hardware inference. As a result, they have become a key technology for enhancing the energy efficiency of artificial intelligence tasks deployed at the edge. 

Mapping SNN architectures to programmable Field-Programmable Gate Arrays (FPGAs) is a critical step in developing custom, scalable inference accelerators. However, translating dense structural connections from software models directly to hardware topologies introduces significant implementation constraints. Specifically, direct spatial mapping often leads to increased interconnect routing pressure, where multi-bit fan-out requirements can exceed available physical wiring tracks. Furthermore, the final classification layer—typically involving combinational \textit{argmax} trees—presents a critical path bottleneck that can severely limit the maximum operating frequency ($F_{\max}$) of the overall hardware system.

To address these challenges, this paper outlines the design, implementation, and verification of a compact, inference-only SNN accelerator optimized for FPGAs. We emphasize that the primary contribution of this work is not a new SNN learning algorithm, but an integrated hardware microarchitecture. Rather than presenting a single optimization in isolation, the principal novelty lies in the holistic integration of time-multiplexed spike distribution, localized weight storage, integer-based LIF processing, and a multi-cycle spike-count and potential-based readout engine. By translating a baseline artificial neural network into this cohesive architecture, this work provides a framework for low-power classification that is verified directly against a software reference model. Specifically, the integrated contributions of this work are:
\begin{enumerate}
    \item A time-multiplexed 1-bit spike-feeding scheme that serializes dense input fan-out into a routable broadcast bus;
    \item A localized memory buffer organization that avoids large, power-intensive combinational weight fan-out;
    \item An integer-based LIF neuron model with register widths sized strictly to avoid overflow;
    \item A multi-cycle pipelined \textit{argmax} and tie-breaker readout engine that removes the dominant combinational critical path; and
    \item A comprehensive timing-closure and power analysis that quantifies the resulting improvements in frequency and energy efficiency.
\end{enumerate}

The remainder of this article is organized as follows: Section~\ref{sec:related_work} reviews the state of the art in neuromorphic hardware. Section~\ref{sec:arch} describes the proposed hardware architecture, including the LIF neuron design and the centralized control flow. Section~\ref{sec:results} details the experimental setup, validation methodology, and implementation results. Finally, Section~\ref{sec:conclusion} concludes the paper.

\section{Related Work}
\label{sec:related_work}
The shift from traditional von Neumann architectures to neuromorphic computing has been driven by the increasing demand for energy-efficient processing at the edge. Spiking Neural Networks (SNNs) have emerged as a prominent solution, leveraging the sparsity and temporal dynamics of biological systems to achieve low-power machine intelligence \cite{roy2019}. While advancements such as the Loihi manycore processor demonstrate the efficiency gains of executing asynchronous, spike-based workloads directly on silicon \cite{loihi2018}, mapping SNN architectures to programmable Field-Programmable Gate Arrays (FPGAs) remains a critical challenge for developing scalable, custom inference accelerators \cite{pourtaherian2018}.

A common paradigm for neuromorphic inference involves converting pre-trained Artificial Neural Networks (ANNs) into SNNs, thereby avoiding the training instability associated with direct spike-based backpropagation \cite{diehl2015}. However, translating dense structural connections from software models to hardware topologies imposes significant physical constraints. Direct spatial mapping on FPGA fabrics often introduces severe interconnect routing pressure, where high-fanout requirements exceed available routing resources and induce substantial combinational propagation delays. To mitigate these bottlenecks, existing literature explores time-multiplexed sequencing, processing-element folding, and centralized finite state machine (FSM) controllers to serialize spike delivery across unified internal buses \cite{neil2014}. These strategies effectively balance parallel execution with design routability, trading absolute single-cycle throughput for improved resource utilization and higher achievable clock frequencies.

Beyond structural routing, the final classification layer of hardware SNN accelerators presents an additional critical path constraint that has received limited attention in literature \cite{perez-carrasco2013}. Traditional hardware implementations of the decision-making stage typically rely on purely combinational \textit{argmax} trees to evaluate winner-take-all behavior within a single clock cycle. As the number of output classes or internal metrics scales, these wide combinational structures introduce significant logic depth, adversely affecting the Worst Negative Slack (WNS) and limiting the maximum operating frequency ($F_{\max}$) of the system. While pipelining is standard practice in conventional deep learning accelerators to isolate long combinational paths, its systematic application within neuromorphic readout units—particularly when managing multi-tier decision rules like spike counting paired with secondary membrane potential tie-breakers—remains less extensively documented \cite{khodabandehloo2018}. Implementing multi-cycle, multi-stage pipelined classification architectures allows for balanced logic depth across successive registers, effectively mitigating timing violations on resource-constrained FPGA fabrics.

\section{Hardware Architecture and Implementation Details}
\label{sec:arch}

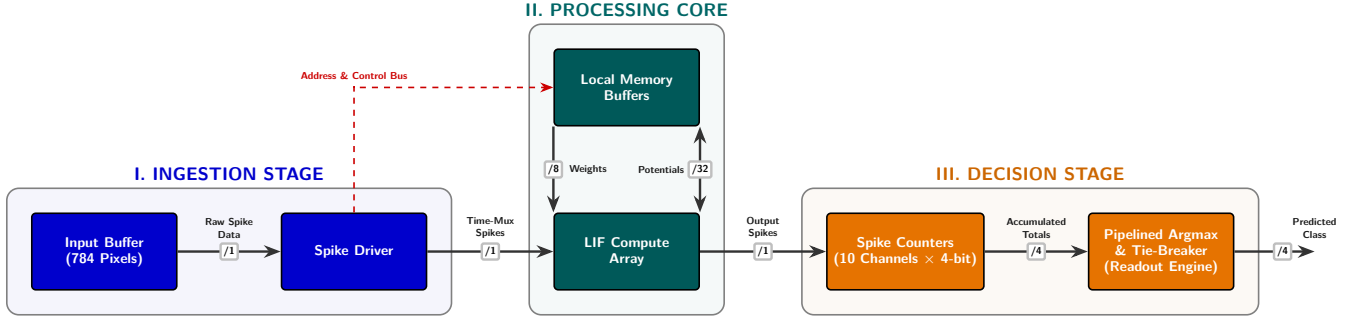
\begin{figure*}
    \centering
    \resizebox{\textwidth}{!}{%
    \begin{tikzpicture}[
        font=\sffamily\scriptsize,
        >=Stealth,
        % ==================== STYLES ====================
        ingest_style/.style={rectangle, draw=black, fill=blue!80!black, text=white, thick, minimum height=3.8em, minimum width=7.2em, align=center, rounded corners=2pt, font=\sffamily\bfseries\scriptsize},
        core_style/.style={rectangle, draw=black, fill=teal!70!black, text=white, thick, minimum height=3.8em, minimum width=7.2em, align=center, rounded corners=2pt, font=\sffamily\bfseries\scriptsize},
        decision_style/.style={rectangle, draw=black, fill=orange!90!black, text=white, thick, minimum height=3.8em, minimum width=7.2em, align=center, rounded corners=2pt, font=\sffamily\bfseries\scriptsize},
        bus/.style={->, very thick, draw=black!80},
        bi_bus/.style={{Stealth[scale=1.0]}-{Stealth[scale=1.0]}, very thick, draw=black!80},
        ctrl/.style={->, thick, dashed, draw=red!80!black},
        bus_tag/.style={midway, fill=white, draw=gray!50, rounded corners=1pt, inner sep=1.5pt, font=\sffamily\tiny\bfseries, text=black}
    ]

    % ==================== STAGE 1: INGESTION STAGE ====================
    \node[ingest_style] (buf) {Input Buffer\\(784 Pixels)};
    \node[ingest_style, right=1.8cm of buf] (driver) {Spike Driver};
    
    % ==================== STAGE 2: PROCESSING CORE ====================
    \node[core_style, right=2.2cm of driver] (lif) {LIF Compute\\Array};
    \node[core_style, above=1.5cm of lif] (bram) {Local Memory\\Buffers};
    
    % ==================== STAGE 3: DECISION STAGE ====================
    \node[decision_style, right=2.2cm of lif] (counters) {Spike Counters\\(10 Channels $\times$ 4-bit)};
    \node[decision_style, right=1.8cm of counters] (argmax) {Pipelined Argmax\\\& Tie-Breaker\\(Readout Engine)};

    % ==================== ROUTING, CONNECTIONS & METRICS ====================
    
    % Ingestion Routing (8-bit raw grayscale pixel data streams)
    \draw[bus] (buf) -- node[above=3pt, font=\sffamily\tiny\bfseries, text=black!90, align=center] {Raw Spike\\Data} 
                         node[bus_tag] {/1} (driver);
    
    % Core Control and Broadcast Bus (1-bit time-multiplexed broadcast)
    \draw[bus] (driver) -- node[above=3pt, font=\sffamily\tiny\bfseries, text=black!90, align=center] {Time-Mux\\Spikes} 
                         node[bus_tag] {/1} (lif);
                          
    % Spike Driver to Local memory
    \draw[ctrl] (driver.north) -- ++(0,2.2) |- 
                          node[pos=0.25, above, font=\sffamily\tiny\bfseries\color{red!80!black}] {Address \& Control Bus} (bram.west);
    
    % Weight Bus: Unidirectional 8-bit signed weight data
    \draw[bus] (bram.south west) -- node[right=4pt, font=\sffamily\tiny\bfseries, text=black!90] {Weights} 
                                      node[bus_tag] {/8} (lif.north west);
                                      
    % Membrane Potential Bus: 32-bit bidirectional state updates to prevent overflow
    \draw[bi_bus] (bram.south east) -- node[left=4pt, font=\sffamily\tiny\bfseries, text=black!90] {Potentials} 
                                          node[bus_tag] {/32} (lif.north east);
    
    % Decision Path Routing
    % 1-bit binary output spike events
    \draw[bus] (lif) -- node[above=3pt, font=\sffamily\tiny\bfseries, text=black!90, align=center] {Output\\Spikes} 
                         node[bus_tag] {/1} (counters);
                         
    % 4-bit parallel counter channels to the Readout Engine
    \draw[bus] (counters) -- node[above=3pt, font=\sffamily\tiny\bfseries, text=black!90, align=center] {Accumulated\\Totals} 
                              node[bus_tag] {/4} (argmax);
                              
    % Final 4-bit BCD/Binary classification output representing digits 0-9
    \draw[bus] (argmax.east) -- ++(1.4,0) node[above=3pt, font=\sffamily\tiny\bfseries, text=black!90, align=center] {Predicted\\Class} 
                                             node[bus_tag, pos=0.6] {/4};

    % ==================== SUB-SYSTEM BOUNDARY BACKGROUNDS ====================
    \begin{scope}[on background layer]
        % Stage I Boundary
        \node[draw=black!50, thick, fill=blue!50!gray!5, rounded corners=6pt, inner sep=12pt, fit=(buf) (driver), 
              label={[font=\sffamily\small\bfseries\color{blue!80!black}]above:I. INGESTION STAGE}] (ingest_group) {};
              
        % Stage II Boundary
        \node[draw=black!50, thick, fill=teal!50!gray!5, rounded corners=6pt, inner sep=12pt, fit=(bram) (lif), 
              label={[font=\sffamily\small\bfseries\color{teal!80!black}]above:II. PROCESSING CORE}] (core_group) {};
              
        % Stage III Boundary
        \node[draw=black!50, thick, fill=orange!50!gray!5, rounded corners=6pt, inner sep=12pt, fit=(counters) (argmax), 
              label={[font=\sffamily\small\bfseries\color{orange!80!black}]above:III. DECISION STAGE}] (decision_group) {};
    \end{scope}

    \end{tikzpicture}
    }%
    \caption{Proposed hardware accelerator microarchitecture optimized for low-cost FPGAs, explicitly illustrating data-path bit-widths, localized dual-bus memory boundaries, and the multi-cycle pipelined decision stage.}
    \label{fig:accelerator_block_diagram}
\end{figure*}

Figure~\ref{fig:accelerator_block_diagram} presents an architectural overview of the proposed SNN accelerator, illustrating the pipelined execution flow structured across three stages: ingestion, processing, and decision-making. At the ingestion stage, incoming stimulus data is held in the Input Buffer and decoded by the Spike Driver into discrete spike events. These events are driven into the Processing Core, where a parallel Leaky Integrate-and-Fire (LIF) Array interfaces with localized memory buffers to fetch and update synaptic weights and membrane potentials. Finally, in the decision stage, output spikes generated by the LIF array are accumulated by counters over a designated temporal window. An Argmax block evaluates these parallel totals to determine the maximum spike activity, outputting the final classification index. This modular, feed-forward topology reduces global routing overhead on the FPGA while supporting parallel processing requirements.

The development of the proposed hardware accelerator begins with establishing an algorithmic baseline in software before translating the design into hardware-compatible primitives. A standard Artificial Neural Network (ANN) featuring a three-layer topology (784-64-10) is trained on the MNIST dataset using the PyTorch library. While a larger configuration containing 128 hidden neurons was initially evaluated, evaluation indicated that a 64-neuron hidden layer achieves a baseline classification accuracy of 97.0\%, which is comparable to the 97.7\% accuracy obtained with the 128-neuron alternative. Reducing the hidden layer size by half reduces the physical routing and logic footprint constraints on the target FPGA fabric during implementation. To avoid the use of floating-point arithmetic units in hardware, the extracted continuous floating-point weights ($W_{f,l}$) for each layer $l$ are mapped to 8-bit signed integers ($W_{q,l}$) through a uniform linear quantization process, defined as:
\begin{equation}
W_{q,l} = \text{clamp}\left( \text{round}\left( S_l \cdot W_{f,l} \right), -128, 127 \right)
\label{eq:weight_quantization}
\end{equation}
where $S_l = 127 / \max(|W_{f,l}|)$ represents a layer-dependent scaling factor computed from the absolute peak weight boundaries of that specific layer, and the clamping operation restricts the outputs to the valid range of an 8-bit signed integer (\textit{int8}) \cite{han2016}. As seen in Equation \ref{eq:weight_quantization}, this quantization methodology allows all subsequent synaptic accumulations within the hardware layers to execute via integer operations, reducing resource consumption and improving operational efficiency in the evaluated setup.

\begin{figure}
    \centering
    \resizebox{\columnwidth}{!}{%
    \begin{tikzpicture}[
        font=\sffamily\scriptsize,
        >=Stealth,
        % ==================== STYLES ====================
        reg/.style={rectangle, draw=black, fill=blue!80!black, text=white, thick, minimum height=0em, minimum width=0em, align=center, rounded corners=2pt, font=\sffamily\bfseries\scriptsize},
        alu/.style={circle, draw=black, fill=red!70!black, text=white, thick, minimum size=0em, align=center, font=\sffamily\bfseries\large},
        comb/.style={rectangle, draw=black, fill=green!60!black, text=white, thick, minimum height=0em, minimum width=0em, align=center, rounded corners=2pt, font=\sffamily\bfseries\scriptsize},
       mux/.style={trapezium, trapezium angle=75, trapezium stretches body, shape border rotate=90, draw=black, fill=orange!90!black, text=white, thick, minimum height=3em, minimum width=4.5em, align=center, font=\sffamily\bfseries\scriptsize},
        comp/.style={circle, draw=black, fill=red!70!black, text=white, thick, minimum size=0em, align=center, font=\sffamily\bfseries\large},
        % Routing Styles
        bus/.style={->, very thick, draw=black!80},
        ctrl/.style={->, thick, dashed, draw=red!80!black},
        bus_tag/.style={midway, fill=white, draw=gray!50, rounded corners=1pt, inner sep=1.5pt, font=\sffamily\tiny\bfseries, text=black}
    ]

    % ==================== COMPONENT PLACEMENT ====================
    % Inputs & Main Pipeline (Y = 0)
    \node[align=right, font=\sffamily\small\bfseries] (isyn_in) at (-3.7, 0) {\textbf{Synaptic Current} \\$I_{syn}$ (32-bit)};

    \node[alu] (add) at (0, 0) {$+$};
    \node[alu] (sub) at (2.5, 0) {$-$};
    \node[reg] (v_reg) at (5, 0) {Membrane\\Potential\\(32-bit)};
    \node[comp] (comp) at (8, 0) {$\ge$};
    \node[reg] (out_reg) at (10.5, 0) {Out FF\\(1-bit)};

    % Threshold & Outputs (Right side)
    \node[align=center, font=\sffamily\scriptsize\bfseries] (threshold) at (8, 1.8) {Threshold ($\theta$)};
    \node[align=left, font=\sffamily\small\bfseries] (spike_out) at (13.8, 0) {Output\\Spike};

    % Feedback Loop Components (Y = -2.5)
    \node[mux] (rst_mux) at (6, -2.5) {RST\\MUX};
    \node[font=\tiny\bfseries, text=white, left=1pt] at ([yshift=1.2em]rst_mux.east) {0};
    \node[font=\tiny\bfseries, text=white, left=1pt] at ([yshift=-1.2em]rst_mux.east) {1};
    
    \node[align=left, font=\sffamily\scriptsize\bfseries, text=gray!80!black] (zero_rst) at (7.5, -2.9) {32'h0};
    \node[comb] (shifter) at (2.5, -2.5) {$\gg leak\_shift$};

    % ==================== DATA BUS ROUTING (Solid Black) ====================
    % Input Path (Direct Current Integration)
    \draw[bus] (isyn_in.east) -- node[bus_tag]{/32} (add.west);

    % Core ALU Pipeline
    \draw[bus] (add.east) -- (sub.west);
    \draw[bus] (sub.east) -- (v_reg.west);
    \draw[bus] (v_reg.east) -- (comp.west);
    \draw[bus] (threshold.south) -- (comp.north);

    % Output Fire Path
    \draw[bus] (comp.east) -- node[above, font=\sffamily\tiny\bfseries]{Fire} (out_reg.west);
    \draw[bus] (out_reg.east) -- node[bus_tag]{/1} (spike_out.west);

    % Feedback Loop: V_reg to RST MUX
    \draw[bus] (v_reg.east) -- ++(1.3,0) |- ([yshift=1.2em]rst_mux.east);
    \draw[bus] (zero_rst.west) |- ([yshift=-1.2em]rst_mux.east);

    % Feedback Loop: RST MUX to Shifter
    \draw[bus] (rst_mux.west) -- ++(-0.6,0) coordinate (rmux_out) -- (shifter.east);
    \draw[bus] (shifter.north) -- (sub.south);

    % Decayed State routing underneath everything (Y = -4.0)
    \draw[bus] (rmux_out) |- (0, -4.0) -- node[pos=-0.05, right, font=\sffamily\tiny\bfseries]{Decayed Base Potential} (add.south);

    % ==================== CONTROL ROUTING (Dashed Red) ====================
    % Feedback Reset Selector
    \path (comp.east) -- (out_reg.west) coordinate[pos=0.5] (fire_tap);
    \draw[ctrl] (fire_tap) -- ++(0, -5.0) coordinate (ctrl_low) -| node[pos=0.25, above, text=red!80!black, font=\tiny\bfseries]{ Feedback Reset Select} (rst_mux.south);

    % ==================== BACKGROUND GROUPING ====================
    \begin{scope}[on background layer]
        % Outer Box: Whole LIF Neuron PE Boundary
        \node[draw=black!60, thick, fill=blue!25!gray!5, rounded corners=8pt, inner sep=20pt, 
              fit=(add) (out_reg) (threshold) (ctrl_low), 
              label={[font=\sffamily\small\bfseries\color{black!80}]above:LIF Neuron Processing Element (PE)}] (pe_box) {};

        % Inner Box: Arithmetic Core
        \node[draw=gray!30, fill=gray!15, rounded corners=4pt, inner sep=12pt, 
              fit=(add) (sub) (shifter), 
              label={[font=\sffamily\small\bfseries, text=gray!60!black]above:Arithmetic Core}] {};
    \end{scope}

    \end{tikzpicture}
    }%
    \caption{Hardware-efficient, multiplier-free RTL microarchitecture of the proposed Processing Element (PE). The architecture implements direct synaptic current ($I_{\text{syn}}$) integration and shift-based leak mechanics mapped directly to the discrete-time dynamics in \eqref{eq:lif_neuron}.}
    \label{fig:pe_microarchitecture}
\end{figure}
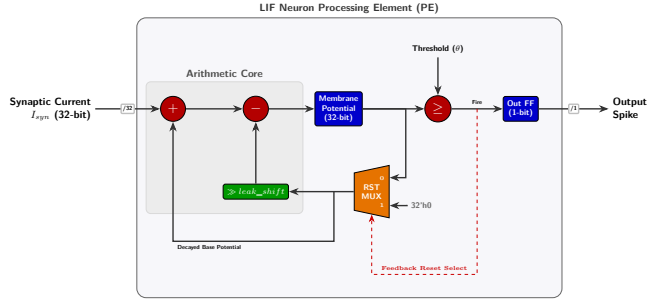

To drive the neuromorphic core, the static 784-pixel MNIST grayscale images are converted into a temporal format compatible with the Leaky Integrate-and-Fire (LIF) neurons. This design utilizes a rate-encoding mechanism where each pixel's normalized intensity determines its firing probability over a discrete temporal window $T$\cite{hunsberger2015}. To implement the Leaky Integrate-and-Fire (LIF) neuron dynamics on the FPGA fabric without requiring hardware multiplication or division units, the continuous-time leakage is realized using an arithmetic right-shift operation. The discrete-time membrane potential $V$ for a given neuron at timestep $t+1$ is modeled as:
\begin{equation}
V[t+1] = V[t] - (V[t] \gg \text{leak\_shift}) + I_{\text{syn}}[t]
\label{eq:lif_neuron}
\end{equation}
where $\text{leak\_shift}$ represents the bit-shift parameter controlling the decay rate, and $I_{\text{syn}}[t]$ is the spatial synaptic current accumulated during the current execution layer. The hardware implementation corresponding to these simplified neuron dynamics is illustrated in Fig.~\ref{fig:pe_microarchitecture}.

To ensure reproducibility, the pseudorandom spike-generation process is evaluated across five independent random seeds (42, 43, 44, 45, and 46) to mitigate potential bias from a single sequence. The length of this temporal processing window, along with the operational parameters of the network, was determined through a parameter grid search evaluating various firing thresholds ($\theta$) and membrane leak rates. To prevent test-set leakage, the 60,000-sample MNIST training dataset was partitioned: the baseline artificial neural network (ANN) was trained on a 59,000-sample split, while a 1,000-sample validation split was used for this SNN parameter grid search. The 10,000-sample MNIST test set was reserved for the final evaluation.

The results of this grid search on the validation split for a temporal window of $T = 16$ are compiled in Table \ref{tab:grid_search_t16}. The evaluation indicates that low thresholds introduce additional temporal noise, whereas higher leak rates (e.g., $\text{leak\_shift} \in \{0, 1\}$) rapidly decay the membrane potential, reducing neural activity and classification accuracy. The grid search shows a maximum validation accuracy of 97.6\% at a threshold of $\theta = 450$, obtained with both $\text{leak\_shift} = 6$ and $\text{leak\_shift} = 7$.

\begin{table}
\caption{SNN Classification Accuracy (\%) Grid Search Results for Time Window $T = 16$ (Evaluated over a 1{,}000-Sample validation Subset)}
\label{tab:grid_search_t16}
\centering
\resizebox{\columnwidth}{!}{%
\begin{tabular}{cccccccll}
\hline
\textbf{Threshold} ($\theta$) & \textbf{sh=0} & \textbf{sh=1} & \textbf{sh=2} & \textbf{sh=3} & \textbf{sh=4} & \textbf{sh=5}  &\textbf{sh=6} &\textbf{sh=7} \\ \hline
25& 68.4& 69.7& 70.3& 70.3& 69.8& 70.1 & 69.8& 70.0\\
50& 76.6& 75.8& 76.2& 75.4& 75.5& 75.5 & 75.4& 75.2\\
100& 88.5& 86.4& 85.1& 84.8& 84.1& 84.2 & 84.3& 84.1\\
200& 92.1& 95.0& 94.3& 94.3& 94.2& 94.1 & 94.0& 94.0\\
300& 90.1& 95.1& 97.1& 96.9& 96.5& 96.6 & 96.4& 96.4\\
350& 87.3& 96.2& 97.1& 97.3& 97.2& 96.9 & 96.8& 96.6\\
400& 83.9& 95.6& 97.4& 97.1& 97.3& 97.5 & 97.1& 97.3\\
\textbf{450}& 78.9& 95.1& 97.2& 97.1& 97.4& 97.4 & \textbf{97.6}& 97.6\\
500& 71.3& 94.1& 97.4& 97.3& 97.0& 97.0 & 97.2& 97.2\\
550& 64.4& 92.9& 97.0& 97.4& 97.2& 96.9 & 97.0& 96.9\\
600& 54.2& 91.5& 97.2& 97.2& 97.2& 97.4 & 97.4& 97.4\\
650& 47.5& 90.7& 96.4& 97.4& 97.4& 97.0 & 97.0& 97.0\\ \hline
\end{tabular}
}%
\end{table}

To resolve this tie, the configuration with $\text{leak\_shift} = 6$ is selected. A higher leak rate reduces sub-threshold membrane accumulation from background noise while still allowing sufficiently driven neurons to cross their activation boundaries. This supports information propagation \cite{thorpe2001, rathi2021} without introducing the noise and random firings observed at lower thresholds (e.g., $\theta \le 200$) \cite{gua2021}. Consequently, the $\theta = 450$ and $\text{leak\_shift} = 6$ configuration is selected as the target configuration for physical VHDL mapping.

Following parameter optimization, the bit-accurate software reference SNN model was evaluated on the 10,000-sample MNIST test set across all five random seeds. Due to extensive simulation runtimes, this full 10,000-sample evaluation was conducted purely in software to evaluate algorithmic variance, while a 1,000-sample subset was utilized to verify the structural equivalency of the VHDL hardware. The results of this software evaluation are detailed in Table \ref{tab:seed_results}. In the evaluated setup, the baseline continuous-activation ANN (trained with Adam for 10 epochs) achieves an evaluated test-set accuracy of 97.70\%. In comparison, the optimized hardware SNN configuration yields a mean test-set accuracy of 96.70\% $\pm$ 0.04\%. This represents an ANN-to-SNN conversion drop of 1.00\%. Given the constraints of the 16-timestep architecture, this result demonstrates that the model preserves the algorithmic fidelity of the baseline network, maintaining stable classification performance across varying temporal noise profiles.

\begin{table}
    \centering
    \caption{Bit-Accurate Software Reference SNN Classification Accuracy on the Full 10,000-Sample MNIST Test Set Across 5 Random Seeds ($\theta = 450$, leak\_shift = 6)}
    \label{tab:seed_results}
    \resizebox{\columnwidth}{!}{%
    \begin{tabular}{cc}
    \hline
    \textbf{Random Seed} & \textbf{Test-Set Accuracy (\%)} \\ \hline
    42& 96.69\\
    43& 96.71\\
    44& 96.65\\
    45& 96.69\\
    46& 96.75\\ \hline
    \textbf{Mean $\pm$ Std Dev}& \textbf{96.70 $\pm$ 0.04}\\ \hline
    \textbf{Baseline ANN (10 epochs, Adam)}& \textbf{97.70}\\ \hline
    \end{tabular}
    }%
\end{table}

\subsection{Centralized Finite State Machine (FSM) and Control Flow}
To coordinate the temporal execution windows and sequential layer dependencies while preventing routing congestion or race conditions, the neuromorphic accelerator is synchronized by a centralized Finite State Machine (FSM) layer controller. This controller manages the progression of algorithmic steps across both structural layers over the duration of the discrete temporal window ($T = 16$). The architectural state progression, as illustrated in Figure \ref{fig:layer_controller_fsm}, is divided into ten deterministic functional states operating in a closed loop.

The functional execution sequence operates via the following structured state progression:
\begin{enumerate}
\item \textbf{IDLE}: The initial quiescent state. The controller holds all internal blocks inactive until an external master handshake trigger asserts the start line (\texttt{start} = '1'), transitioning the FSM into the hidden layer loading cycle.

\item \textbf{LOAD\_HIDDEN\_LAYER}: A 784-clock-cycle iteration over the input pixel space ($0 \le \text{pixel\_counter} \le 783$). At each clock edge, the \texttt{input\_spike\_driver} broadcasts a single 1-bit pixel spike across a shared bus to all 64 hidden layer synapse accumulators simultaneously. Each accumulator concurrently indexes its localized memory to aggregate the respective 8-bit weight if a spike is present.

\begin{figure}
    \centering
    \resizebox{\columnwidth}{!}{%
    \begin{tikzpicture}[
        font=\sffamily\scriptsize,
        >=Stealth,
        node distance=1.0cm and 2.0cm,
        % ==================== STYLES ====================
        % Idle State
        state_idle/.style={
            rectangle, draw=black, fill=green!60!black, text=white, very thick, 
            minimum height=3.2em, minimum width=18em, align=center, rounded corners=4pt, 
            font=\sffamily\scriptsize
        },
        % Loading Loops
        state_load/.style={
            rectangle, draw=black, fill=orange!90!black, text=white, very thick, 
            minimum height=3.2em, minimum width=18em, align=center, rounded corners=4pt, 
            font=\sffamily\scriptsize
        },
        % Core Sync & Processing States
        state_process/.style={
            rectangle, draw=black, fill=blue!80!black, text=white, very thick, 
            minimum height=3.2em, minimum width=18em, align=center, rounded corners=4pt, 
            font=\sffamily\scriptsize
        },
        % Readout & Evaluation States
        state_eval/.style={
            rectangle, draw=black, fill=violet!70!black, text=white, very thick, 
            minimum height=3.2em, minimum width=18em, align=center, rounded corners=4pt, 
            font=\sffamily\scriptsize
        },
        % Cleanup / Reset Stat
        state_reset/.style={
            rectangle, draw=black, fill=red!70!black, text=white, very thick, 
            minimum height=3.2em, minimum width=18em, align=center, rounded corners=4pt, 
            font=\sffamily\scriptsize
        },
        % Routing & Condition Tag Styles
        bus/.style={->, very thick, draw=black!80},
        bus_tag/.style={midway, fill=white, draw=gray!50, rounded corners=1pt, inner sep=2.5pt, font=\sffamily\tiny\bfseries, text=black, align=center}
    ]
    
    % ==================== STATE NODE PLACEMENT ====================
    \node[state_idle] (idle) {\textbf{IDLE STATE}\\Initial quiescent state};
    
    \node[state_load, below=of idle] (load_hidden) {\textbf{LOAD\_HIDDEN\_LAYER}\\Loop: pixel\_counter 0 to 783};
    
    \node[state_process, below=of load_hidden] (hidden_isyn) {\textbf{HIDDEN\_ISYN\_READY}\\Assert prep\_Isyn\_hidden};
    
    \node[state_process, below=of hidden_isyn] (hidden_lif) {\textbf{HIDDEN\_LIF\_PROCESSING}\\Assert en\_lif\_hidden};
    
    \node[state_load, below=of hidden_lif] (load_output) {\textbf{LOAD\_OUTPUT\_LAYER}\\Loop: neuron\_counter 0 to 63};
    
    \node[state_process, below=of load_output] (output_isyn) {\textbf{OUTPUT\_ISYN\_READY}\\Assert prep\_Isyn\_output};
    
    \node[state_process, below=of output_isyn] (output_lif) {\textbf{OUTPUT\_LIF\_PROCESSING}\\Assert en\_lif\_output};
    
    \node[state_eval, below=of output_lif] (output_spike) {\textbf{OUTPUT\_SPIKE\_COUNTING}\\Assert count\_spike\_valid};
    
    \node[state_eval, below=of output_spike] (argmax) {\textbf{ARGMAX\_COMPUTING}\\\textbf{Assert} calculate\_argmax \ $\vert$ \ \textbf{Wait}: prediction\_ready = '1'};
    
    \node[state_reset, below=of argmax] (reset) {\textbf{RESET\_STATE}\\Clear reset\_counter, rst\_V};

    % ==================== FORWARD PATH ROUTING ====================
    \draw[bus] (idle) -- node[bus_tag] {start = '1'} (load_hidden);
    
    \draw[bus] (load_hidden) -- node[bus_tag] {pixel\_counter == 783} (hidden_isyn);
    
    \draw[bus] (hidden_isyn) -- (hidden_lif);
    
    \draw[bus] (hidden_lif) -- (load_output);
    
    \draw[bus] (load_output) -- node[bus_tag] {neuron\_counter == 63} (output_isyn);
    
    \draw[bus] (output_isyn) -- (output_lif);
    
    \draw[bus] (output_lif) -- (output_spike);
    
    \draw[bus] (output_spike) -- node[bus_tag] {timestep\_counter == T - 1} (argmax);
    
    \draw[bus] (argmax) -- node[bus_tag] {sample\_counter $<$ NUM\_SAMPLES - 1} (reset);

    % ==================== FEEDBACK LOOPS & MARGIN ROUTING (ROUNDED) ====================
    
    % Right Margin Loop 1: Hidden Layer Pixel Iteration
    \coordinate (lh_out) at ([yshift=-10pt]load_hidden.east);
    \coordinate (lh_in) at ([yshift=10pt]load_hidden.east);
    \draw[bus, rounded corners=4pt] (lh_out) -- ++(1.3,0) |- node[bus_tag, pos=0.25] {pixel\_counter $<$ 783} (lh_in);
    
    % Right Margin Loop 2: Output Layer Neuron Iteration 
    \coordinate (lo_out) at ([yshift=-10pt]load_output.east);
    \coordinate (lo_in) at ([yshift=10pt]load_output.east);
    \draw[bus, rounded corners=4pt] (lo_out) -- ++(1.3,0) |- node[bus_tag, pos=0.25] {neuron\_counter $<$ 63} (lo_in);
    
    % Left Margin Loop 1: Timestep Pipeline Return
    \draw[bus, rounded corners=4pt] (output_spike.west) -- ++(-1.4,0) |- node[bus_tag, pos=0.25] {timestep\_counter $<$ T - 1\\Increment timestep} ([yshift=-8pt]load_hidden.west);
    
    % Left Margin Loop 2: Reset to Next Sample
    \draw[bus, rounded corners=4pt] (reset.west) -- ++(-3,0) |- node[bus_tag, pos=0.25] {Next Sample} ([yshift=8pt]load_hidden.west);
    
    % Right Margin Loop 3: Complete Execution Back to Idle 
    \draw[bus, rounded corners=4pt] (argmax.east) -- ++(2.5,0) |- node[bus_tag, pos=0.1] {sample\_counter ==\\NUM\_SAMPLES - 1\\(Final Sample)} (idle.east);
    
    \end{tikzpicture}
    }%
    \caption{Algorithmic state machine diagram of the centralized FSM layer controller.}
    \label{fig:layer_controller_fsm}
\end{figure}
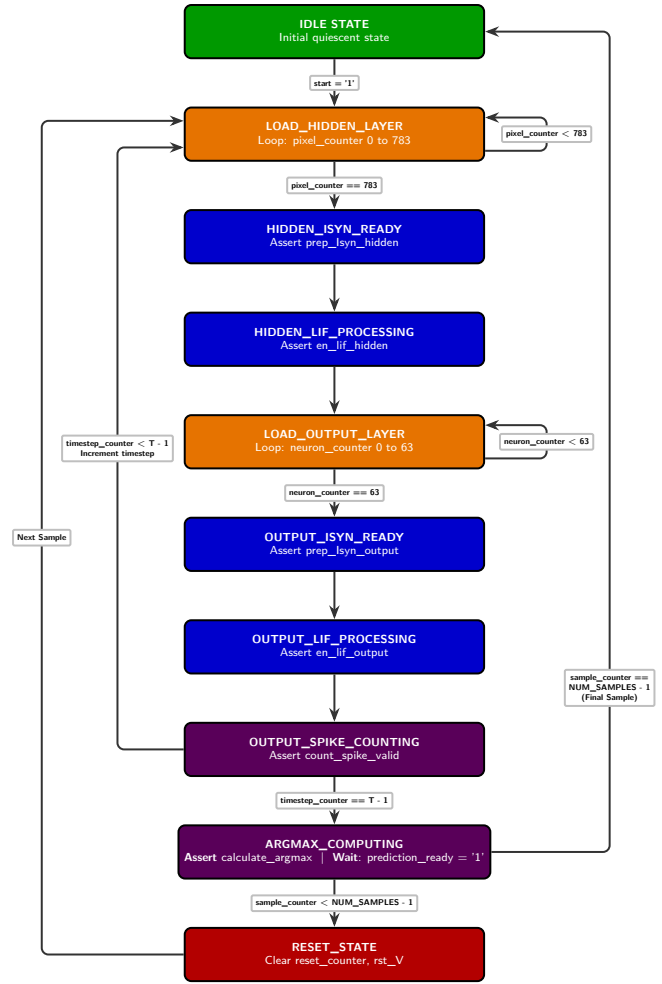

\item \textbf{HIDDEN\_ISYN\_READY}: A single-cycle synchronization barrier reached immediately after the 784-pixel iteration. The controller asserts \texttt{prep\_Isyn\_hidden}, causing all 64 hidden accumulators to latch their total internal current sums onto the 32-bit \texttt{I\_syn} ports of the corresponding hidden LIF neurons while clearing internal tracking registers for the next timestep.

\item \textbf{HIDDEN\_LIF\_PROCESSING}: A single-cycle state executing parallel neuron evaluation. Asserting \texttt{en\_lif\_hidden} = '1' activates all 64 hidden Leaky Integrate-and-Fire (LIF) neurons concurrently. On the rising clock edge, each neuron updates its membrane potential via a multiplierless right-shift leak subtraction and synaptic current addition (Equation \ref{eq:lif_neuron}). Neurons exceeding the threshold ($\theta$) register an output spike ('1') and reset their membrane potential register ($V$) to zero; otherwise, the spike register is cleared ('0') and $V_{\text{next}}$ is retained.

\item \textbf{LOAD\_OUTPUT\_LAYER}: A 64-clock-cycle iteration over the hidden neuron layer ($0 \le \text{neuron\_}\allowbreak\text{counter} \le 63$). At each rising clock edge, the registered 1-bit output spike from the active hidden neuron is broadcast to all 10 classification output-layer synapse accumulators in parallel. Each accumulator concurrently samples this line and indexes its local weight inside its internal register.

\item \textbf{OUTPUT\_ISYN\_READY}: A single-cycle memory barrier following the 64-neuron iteration. The controller asserts \texttt{prep\_Isyn\_output} to latch the 10 accumulated synaptic current values directly onto the 32-bit \texttt{I\_syn} input lines of the output layer LIF neurons, while resetting the synapse accumulator blocks and weight pointers back to zero.

\item \textbf{OUTPUT\_LIF\_PROCESSING}: A single-cycle execution state where the controller asserts \texttt{en\_}\allowbreak\texttt{lif\_}\allowbreak\texttt{output} = '1'. This drives the 10 output-layer LIF neurons to compute their updated potentials and evaluate threshold boundaries concurrently.

\item \textbf{OUTPUT\_SPIKE\_COUNTING}: A single-cycle evaluation state where the controller asserts \texttt{count\_}\allowbreak\texttt{spike\_}\allowbreak\texttt{valid}, instructing 10 independent tracking registers to sample the output spike lines and increment if a firing event occurred. If the temporal window is incomplete ($\text{timestep\_counter} < T - 1$), the loop increments and transitions back to \textbf{LOAD\_HIDDEN\_LAYER}; if completed, it advances to readout.

\item \textbf{ARGMAX\_COMPUTING}: Reached at the end of the temporal processing window ($t = T-1$). The controller asserts \texttt{calculate\_argmax} and waits until the argmax unit asserts \texttt{prediction\_ready} = '1'. If more validation images remain, the FSM triggers a system reset via \texttt{reset\_counter} and \texttt{rst\_V} and advances to the \textbf{RESET\_STATE}; if the final sample is reached, it returns to \textbf{IDLE}.

\item \textbf{RESET\_STATE}: Positioned at the end of a complete sample execution window. Over one clock cycle, the controller drives the clear lines back to '0' after clearing prior membrane accumulations and spike registers across the fabric, before transitioning back into \textbf{LOAD\_HIDDEN\_LAYER} to process the first timestep of the subsequent sample.
\end{enumerate}

\subsection{Memory Architecture and Time-Multiplexed Spike Broadcasting}
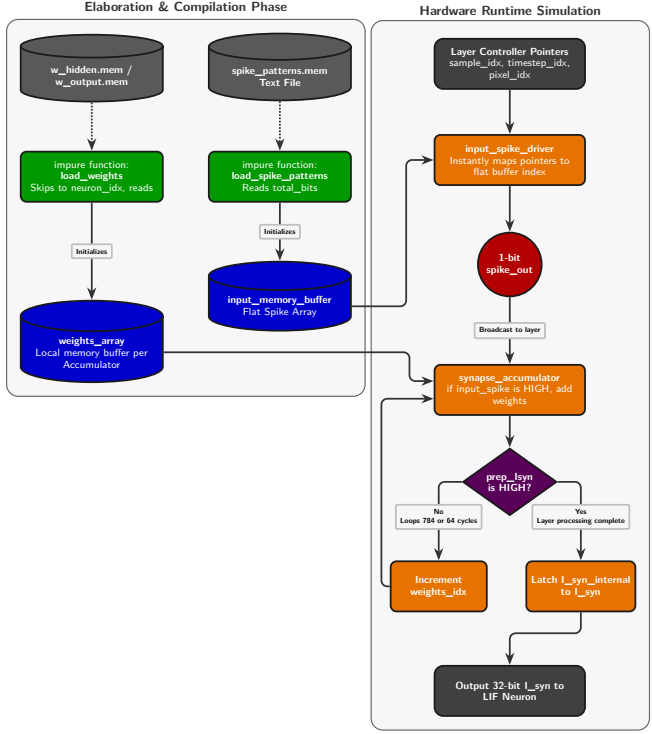
\begin{figure}
    \centering
    \resizebox{\columnwidth}{!}{%
    \begin{tikzpicture}[
        font=\sffamily\scriptsize,
        >=Stealth,
        % ==================== STYLES ====================
        database/.style={cylinder, shape border rotate=90, draw=black!90, very thick, aspect=0.25, align=center, minimum width=3.4cm, minimum height=1.6cm, fill=gray!70!black, text=white},
        blue_database/.style={cylinder, shape border rotate=90, draw=black!90, very thick, aspect=0.25, align=center, minimum width=3.4cm, minimum height=1.6cm, fill=blue!80!black, text=white},
        green_func/.style={rectangle, rounded corners=4pt, draw=black!90, very thick, align=center, minimum width=3.4cm, minimum height=1.2cm, fill=green!60!black, text=white},
        yellow_box/.style={rectangle, rounded corners=4pt, draw=black!90, very thick, align=center, minimum width=3.6cm, minimum height=1.2cm, fill=orange!90!black, text=white},
        grey_box/.style={rectangle, rounded corners=6pt, draw=black!90, very thick, align=center, minimum width=3.6cm, minimum height=1.2cm, fill=gray!50!black, text=white},
        diamond_dec/.style={diamond, aspect=1.4, draw=black!90, very thick, align=center, fill=violet!70!black, text=white, inner sep=3pt},
        red_circle/.style={circle, draw=black!90, very thick, align=center, minimum size=1.5cm, fill=red!70!black, text=white},
        tag/.style={fill=blue!25!gray!5, draw=gray!50, rounded corners=1pt, inner sep=3pt, font=\sffamily\tiny\bfseries, text=black, align=center},
        arrow/.style={->, very thick, draw=black!80},
        dotted_arrow/.style={->, very thick, draw=black!80, densely dotted}
    ]
    
    % ==================== PHASE 1 NODES ====================
    \node[database] (w_mem) at (0, -0.3) {\textbf{w\_hidden.mem /}\\\textbf{w\_output.mem}};
    \node[database] (spike_mem) at (4.5, -0.3) {\textbf{spike\_patterns.mem}\\\textbf{Text File}};
    \node[green_func] (load_w) at (0, -2.7) {impure function:\\\textbf{load\_weights}\\Skips to neuron\_idx, reads};
    \node[green_func] (load_s) at (4.5, -2.7) {impure function:\\\textbf{load\_spike\_patterns}\\Reads total\_bits};
    \node[blue_database] (in_buf) at (4.5, -5.8) {\textbf{input\_memory\_buffer}\\Flat Spike Array};
    \node[blue_database] (w_arr) at (0, -6.9) {\textbf{weights\_array}\\Local memory buffer per\\Accumulator};

    % ==================== PHASE 2 NODES ====================
    \node[grey_box] (pointers) at (10, 0) {\textbf{Layer Controller Pointers}\\sample\_idx, timestep\_idx,\\pixel\_idx};
    \node[yellow_box] (driver) at (10, -2.3) {\textbf{input\_spike\_driver}\\Instantly maps pointers to\\flat buffer index};
    \node[red_circle] (spike_out) at (10, -4.8) {\textbf{1-bit}\\\textbf{spike\_out}};
    \node[yellow_box] (acc) at (10, -7.8) {\textbf{synapse\_accumulator}\\if input\_spike is HIGH, add\\weights};
    \node[diamond_dec] (decision) at (10, -10) {\textbf{prep\_Isyn}\\\textbf{is HIGH?}};
    \node[yellow_box, minimum width=2.3cm] (inc) at (8.3, -12.5) {\textbf{Increment}\\\textbf{weights\_idx}};
    \node[yellow_box, minimum width=2.3cm] (latch) at (11.7, -12.5) {\textbf{Latch I\_syn\_internal}\\\textbf{to I\_syn}};
    \node[grey_box] (out_neuron) at (10, -15.0) {\textbf{Output 32-bit I\_syn to}\\\textbf{LIF Neuron}};

    % ==================== INTRA-PHASE ROUTING ====================
    % Phase 1 Routing
    \draw[dotted_arrow] (w_mem) -- (load_w);
    \draw[dotted_arrow] (spike_mem) -- (load_s);
    \draw[arrow] (load_w) -- node[tag] {Initializes} (w_arr);
    \draw[arrow] (load_s) -- node[tag] {Initializes} (in_buf);

    % Phase 2 Vertical Flow
    \draw[arrow] (pointers) -- (driver);
    \draw[arrow] (driver) -- (spike_out);
    \draw[arrow] (spike_out) -- node[tag] {Broadcast to layer} (acc);
    \draw[arrow] (acc) -- (decision);
    
    % Condition branches 
    \draw[arrow, rounded corners=4pt] (decision.west) -| node[tag, pos=0.5, below, yshift=-15pt] {No\\Loops 784 or 64 cycles} (inc.north);
    \draw[arrow, rounded corners=4pt] (decision.east) -| node[tag, pos=0.5, below, yshift=-15pt] {Yes\\Layer processing complete} (latch.north);
    
    \draw[arrow, rounded corners=4pt] (latch.south) -- ++(0,-0.5)  -| (out_neuron.north);
    
    % Loop back routing line (Rounded Corners Added)
    \draw[arrow, rounded corners=4pt] (inc.west) -- ++(-0.2,0) |- ([yshift=-6pt]acc.west);

    % ==================== CROSS-PHASE ROUTING (ROUNDED) ====================
    % From input memory to spike driver
    \draw[arrow, rounded corners=4pt] (in_buf.east) -- ++(1.3,0) |- (driver.west);
    
    % From weights array to accumulator
    \draw[arrow, rounded corners=4pt] (w_arr.east) -- (7.6, -6.9) |- ([yshift=6pt]acc.west);

    % ==================== BACKGROUND GROUPING ====================
    \begin{scope}[on background layer]
        % Top alignment anchor
        \coordinate (phase2_top_align) at (7, 0 |- w_mem.north);
        
        % Outer Box: Elaboration & Compilation Phase
        \node[draw=black!60, thick, fill=blue!25!gray!5, rounded corners=8pt, inner sep=9pt, 
              fit=(w_mem) (spike_mem) (load_w) (load_s) (in_buf) (w_arr), 
              label={[font=\sffamily\small\bfseries\color{black!80}]above:Elaboration \& Compilation Phase}] (phase1_box) {};

        % Outer Box: Hardware Runtime Simulation
        \node[draw=black!60, thick, fill=blue!25!gray!5, rounded corners=8pt, inner sep=9pt, 
              fit=(pointers) (driver) (spike_out) (acc) (decision) (inc) (latch) (out_neuron) (phase2_top_align), 
              label={[font=\sffamily\small\bfseries\color{black!80}]above:Hardware Runtime Simulation}] (phase2_box) {};
    \end{scope}

\end{tikzpicture}
}%
\caption{Structural breakdown of compilation-level elaboration versus physical hardware runtime execution.}
\label{fig:input_spike_driver_mem}
\end{figure}

Directly mapping a dense 784-to-64 layer onto an FPGA fabric can result in high interconnect fan-out, requiring over 400,000 signal lines (784 $\times$ 64 $\times$ 8-bit weights), potentially leading to routing congestion and increased critical-path delays. To mitigate these routing constraints, the proposed architecture combines a time-multiplexed 1-bit broadcasting bus with an isolated, localized buffer memory organization scheme. This approach splits the implementation into distinct synthesis-time phases and runtime execution loops, as illustrated in Figure \ref{fig:input_spike_driver_mem}.

During design elaboration and hardware compilation, the synthesizer removes the need for external runtime flash communication or high-overhead memory interfaces by executing VHDL compile-time \textit{impure functions}. The function \texttt{load\_spike\_patterns} opens raw, formatted text data vectors and sequentially flattens the complete multi-sample dataset directly into a localized, un-routed continuous hardware array known as the \texttt{input\_memory\_buffer}. Simultaneously, the function \texttt{load\_weights} processes the quantized 8-bit integer matrix elements, parsing and distributing them directly into local look-up structures dedicated to each individual synapse accumulator module.

Operating in parallel, each of the 64 synapse accumulators monitors the state of the unified 1-bit broadcast line. At each clock edge, the accumulator modules use the global \texttt{pixel\_idx} pointer to address their own local buffer, retrieving the specific 8-bit signed weight associated with the current pixel sequence. If the broadcasted input spike bit is evaluated as \texttt{HIGH} ('1'), the accumulator adds the retrieved weight value to an internal 32-bit accumulation register. Conversely, if the spike line is \texttt{LOW} ('0'), the accumulation stage is bypassed, reducing dynamic power consumption.

This architecture converts a spatial routing bottleneck into a time-multiplexed sequence. Instead of utilizing extensive routing resources and logic adders within a single clock cycle, the network processes the connections over 784 consecutive clock cycles. Upon completion of the loop, the accumulated sums are latched onto the 32-bit \texttt{I\_syn} output bus feeding the LIF neurons, and the internal accumulation registers are cleared to prepare for the next evaluation cycle.

\subsection{Hardware Verification and Automated Grading Testbench}
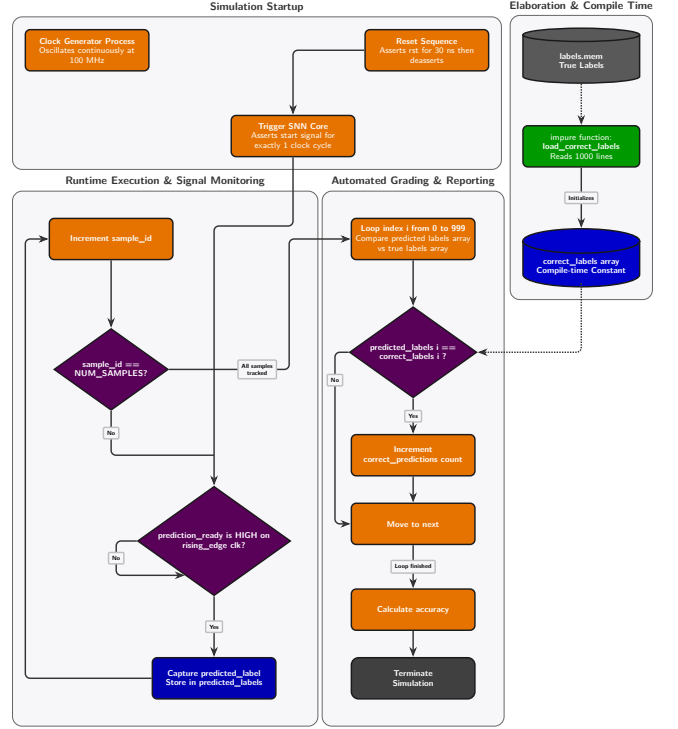
\begin{figure}
    \centering
    \resizebox{\columnwidth}{!}{%
    \begin{tikzpicture}[
        font=\sffamily\scriptsize,
        >=Stealth,
        % ==================== STYLES ====================
        database/.style={cylinder, shape border rotate=90, draw=black!90, very thick, aspect=0.25, align=center, minimum width=3.4cm, minimum height=1.6cm, fill=gray!70!black, text=white},
        blue_database/.style={cylinder, shape border rotate=90, draw=black!90, very thick, aspect=0.25, align=center, minimum width=3.4cm, minimum height=1.6cm, fill=blue!80!black, text=white},
        green_func/.style={rectangle, rounded corners=4pt, draw=black!90, very thick, align=center, minimum width=3.4cm, minimum height=1.2cm, fill=green!60!black, text=white},
        yellow_box/.style={rectangle, rounded corners=4pt, draw=black!90, very thick, align=center, minimum width=3.6cm, minimum height=1.2cm, fill=orange!90!black, text=white},
        blue_box/.style={rectangle, rounded corners=4pt, draw=black!90, very thick, align=center, minimum width=3.6cm, minimum height=1.2cm, fill=blue!70!black, text=white},
        grey_box/.style={rectangle, rounded corners=6pt, draw=black!90, very thick, align=center, minimum width=3.6cm, minimum height=1.2cm, fill=gray!50!black, text=white},
        diamond_dec/.style={diamond, aspect=1.4, draw=black!90, very thick, align=center, fill=violet!70!black, text=white, inner sep=3pt},
        tag/.style={fill=blue!25!gray!5, draw=gray!50, rounded corners=1pt, inner sep=3pt, font=\sffamily\tiny\bfseries, text=black, align=center},
        arrow/.style={->, very thick, draw=black!80},
        dotted_arrow/.style={->, very thick, draw=black!80, densely dotted}
    ]
    
    % ==================== Elaboration & Compile Time ====================
    \node[database] (labels_mem) at (10.7, 7.0) {\textbf{labels.mem}\\\textbf{True Labels}};
    \node[green_func] (load_labels) at (10.7, 4.5) {impure function:\\\textbf{load\_correct\_labels}\\Reads 1000 lines};
    \node[blue_database] (correct_arr) at (10.7, 1.0) {\textbf{correct\_labels array}\\\textbf{Compile-time Constant}};

    % ==================== Simulation Startup ====================
    \node[yellow_box] (clk_gen) at (-3.7, 7.3) {\textbf{Clock Generator Process}\\Oscillates continuously at\\100 MHz};
    \node[yellow_box] (reset_seq) at (6.2, 7.3) {\textbf{Reset Sequence}\\Asserts rst for 30 ns then\\deasserts};
    \node[yellow_box] (trigger_snn) at (2.3, 4.8) {\textbf{Trigger SNN Core}\\Asserts start signal for\\exactly 1 clock cycle};

    % ==================== Runtime Execution & Signal Monitoring  ====================
    \node[yellow_box] (inc_sample) at (-3.0, 1.8) {\textbf{Increment sample\_id}};
    \node[diamond_dec] (sample_cond) at (-3.0, -2) {\textbf{sample\_id ==}\\\textbf{NUM\_SAMPLES?}};
    \node[diamond_dec] (pred_ready) at (0.0, -7) {\textbf{prediction\_ready is HIGH on}\\\textbf{rising\_edge clk?}};
    \node[blue_box] (capture_label) at (0.0, -11) {\textbf{Capture predicted\_label}\\\textbf{Store in predicted\_labels}};

    % ==================== Automated Grading & Reporting  ====================
    \node[yellow_box] (loop_index) at (5.8, 1.8) {\textbf{Loop index i from 0 to 999}\\Compare predicted labels array \\vs true labels array};
    \node[diamond_dec] (pred_equal) at (5.8, -1.5) {\textbf{predicted\_labels i ==}\\\textbf{correct\_labels i ?}};
    \node[yellow_box] (inc_correct) at (5.8, -4.5) {\textbf{Increment}\\\textbf{correct\_predictions count}};
    \node[yellow_box] (move_next) at (5.8, -6.5) {\textbf{Move to next}};
    \node[yellow_box] (calc_acc) at (5.8, -9.0) {\textbf{Calculate accuracy}};
    \node[grey_box] (terminate) at (5.8, -11.0) {\textbf{Terminate}\\\textbf{Simulation}};

    % ==================== ALIGNMENT ANCHORS ====================
    \coordinate (p1_top) at (11.0, 7.2);
    \coordinate (p2_top) at (3.25, 7.2);
    \coordinate (p3_top) at (-5.1, 2.2); 
    \coordinate (p4_top) at (5.5, 2.2);

    % ==================== ROUTING & CONNECTIONS ====================
    % Phase 1 Routing 
    \draw[dotted_arrow] (labels_mem) -- (load_labels);
    \draw[arrow] (load_labels) -- node[tag] {Initializes} (correct_arr);
    \draw[dotted_arrow, rounded corners=4pt] (correct_arr.south) -- ++(-0.0,0) |- ([yshift=0pt]pred_equal.east);

    % Phase 2 Routing
    \draw[arrow, rounded corners=4pt] (reset_seq) -| (trigger_snn);
    \draw[arrow, rounded corners=4pt] (trigger_snn.south) -- ++(0,-2.0) -| (pred_ready.north);

    % Phase 3 Routing 
    \draw[arrow] (inc_sample) -- (sample_cond);
    \draw[very thick, draw=black!80, rounded corners=4pt] (sample_cond.south) |- node[tag, pos=0.25] {No} (0.0, -4.5);
    \draw[arrow, rounded corners=4pt] (pred_ready.west) -- ++(-0.6,0) |- node[tag, midway, pos=0.25] {No} ([xshift=1.4cm,yshift=-1cm]pred_ready.west);
    \draw[arrow] (pred_ready) -- node[tag] {Yes} (capture_label);
    \draw[arrow, rounded corners=4pt] (capture_label.west) -- (-5.5, -11) |- (inc_sample.west);

    % Cross Phase 3 to Phase 4 
    \draw[arrow, rounded corners=4pt] (sample_cond.east) -- node[tag, left, pos=0.9] {All samples\\ tracked} ++(3.5,0.0) |- (loop_index.west);

    % Phase 4 Routing 
    \draw[arrow] (loop_index) -- (pred_equal);
    \draw[arrow] (pred_equal) -- node[tag, midway, pos=0.5] {Yes} (inc_correct);
    \draw[arrow, rounded corners=4pt] (pred_equal.west) -- ++(-0.4,0) |- node[tag, pos=0.1, above] {No}  (move_next.west);
    \draw[arrow] (inc_correct) -- (move_next);
    \draw[arrow] (move_next) -- node[tag, midway] {Loop finished} (calc_acc);
    \draw[arrow] (calc_acc) -- (terminate);

    % ==================== BACKGROUND BOUNDING BOXES ====================
    \begin{scope}[on background layer]
        % 1. Elaboration & Compile Time
        \node[draw=black!60, thick, fill=blue!25!gray!5, rounded corners=8pt, inner sep=10pt, 
              fit=(labels_mem) (load_labels) (correct_arr) (p1_top), 
              label={[font=\sffamily\small\bfseries\color{black!80}]above:Elaboration \& Compile Time}] (phase1_box) {};

        % 2. Simulation Startup
        \node[draw=black!60, thick, fill=blue!25!gray!5, rounded corners=8pt, inner sep=10pt, 
              fit=(clk_gen) (reset_seq) (trigger_snn) (p2_top), 
              label={[font=\sffamily\small\bfseries\color{black!80}]above:Simulation Startup}] (phase2_box) {};

        % 3. Runtime Execution & Signal Monitoring 
        \node[draw=black!60, thick, fill=blue!25!gray!5, rounded corners=8pt, inner sep=22pt, 
              fit=(inc_sample) (sample_cond) (pred_ready) (capture_label) (p3_top), 
              label={[font=\sffamily\small\bfseries\color{black!80}]above:Runtime Execution \& Signal Monitoring}] (phase3_box) {};

        % 4. Automated Grading & Reporting 
        \node[draw=black!60, thick, fill=blue!25!gray!5, rounded corners=8pt, inner sep=22pt, 
              fit=(loop_index) (pred_equal) (inc_correct) (move_next) (calc_acc) (terminate) (p4_top), 
              label={[font=\sffamily\small\bfseries\color{black!80}]above:Automated Grading \& Reporting}] (phase4_box) {};
    \end{scope}

\end{tikzpicture}
}%
\caption{Architectural flowchart of the automated VHDL testbench, detailing the hardware runtime simulation, synchronous handshaking, and the automated post-simulation grading loop.}
\label{fig:testbench_flow}
\end{figure}

To assess functional equivalence between the sequential Python reference model and the synthesized VHDL architecture in the tested simulation setup, an automated, self-checking testbench was developed. Rather than manually inspecting timing waveforms for individual inference cycles, the testbench environment was designed to execute a continuous, batch-processing VHDL simulation of $1,000$ samples from the MNIST test set, automatically evaluating the hardware's predictive accuracy against the software's reference baseline. The structural flow of this verification environment is illustrated in Fig. \ref{fig:testbench_flow}.

The simulation environment is driven by a $100\text{ MHz}$ testbench clock generator. The simulation begins with a $30\text{ ns}$ global reset sequence to initialize all internal neuron membrane registers, state machines, and synapse accumulators. Following the reset, the testbench increments a master sample counter ($0$ to $999$) and triggers the top-level SNN module by asserting the \texttt{start} signal for a single clock cycle.

The testbench suspends its stimulus generation and waits for the SNN core to assert the \texttt{prediction\allowbreak\_ready} flag. Upon detecting the rising edge of this flag, the testbench captures the 4-bit output on the \texttt{predicted\_label} bus and logs it into a localized memory array. This handshake sequence is repeated until all $1,000$ images have been processed.

Once the hardware runtime simulation concludes, the testbench enters an automated verification phase. The testbench iterates through the stored array of hardware predictions, comparing each entry against a pre-loaded ground-truth memory array containing the true MNIST labels. The simulation console tallies the correct predictions and calculates the final inference accuracy, demonstrating that the integer-based hardware aligns with the Python baseline in the evaluated setup without requiring manual waveform inspection.

\section{Experimental Results and Hardware Evaluation}
\label{sec:results}
To establish a transparent and rigorous evaluation framework, the experimental analysis in this work is structured across three distinct scopes: (1) algorithmic fidelity and variance are evaluated using the 10,000-sample, bit-accurate software SNN reference model across five random seeds; (2) functional equivalence of the VHDL hardware is validated against this software baseline using a 1,000-sample simulation testbench; and (3) physical hardware performance metrics---including operating frequency, processing latency, and on-chip power consumption---are derived exclusively from Vivado post-implementation timing and power analyses.

To evaluate the feasibility, timing performance, and resource efficiency of the proposed Spiking Neural Network (SNN) hardware accelerator, synthesis and implementation were conducted using the AMD Vivado Design Suite targeting the AMD Artix-7 AC701 Evaluation Platform XC7A200T FPGA. As an entry-level, low-power FPGA platform, this target device serves as a benchmark for evaluating the performance and resource efficiency of the microarchitecture in the evaluated setup.

To establish a comparative baseline, the high-level Python algorithmic reference model was profiled sequentially on a host system equipped with an Intel Core Ultra 9 275HX processor operating at a maximum turbo frequency of 5.4 GHz. This evaluation framework enables a comparison between sequential software execution and the parallelized FPGA implementations. Consequently, this section presents an empirical performance analysis benchmarking the initial non-pipelined readout architecture (Version 1) and the multi-cycle pipelined sequential argmax core (Version 2) against the CPU software reference. Both hardware configurations are evaluated under timing constraints defined within a hardware constraints file (\texttt{constants.xdc}) to characterize critical path propagation delays, Worst Negative Slack (WNS), estimated maximum operating frequencies ($F_{max}$), estimated on-chip power dissipation, and Look-Up Table (LUT) fabric utilization. To support experimental reproducibility, the implementation details, synthesis configurations, and system-level parameters for both design versions are summarized in Table~\ref{tab:reproducibility}.

\begin{table}
\caption{Reproducibility and Implementation Details.}
\label{tab:reproducibility}
\centering
\resizebox{\columnwidth}{!}{%
\begin{threeparttable}
\begin{tabular}{ll}
\toprule
\textbf{Item} & \textbf{Value} \\
\midrule
FPGA board / evaluation platform & AMD Artix-7 AC701\\
Exact FPGA part & xc7a200tfbg676-2\\
Vivado version & 2023.2\\
Synthesis strategy & Vivado Synthesis Defaults\\
Implementation strategy & V1: Vivado Implementation Defaults\\
                        & V2: Performance\_ExtraTimingOpt\\
Target clock constraint (V1 / V2) & 75~ns / 6~ns \\
Achieved $F_{max}$ (V1 / V2) & 13.3~MHz / 167~MHz \\
LUTs (abs.\ / \%) & V1: 17,021 (12.72\%) out of 133,800\\
                  & V2: 15,985 (11.95\%) out of 133,800\\
Flip-Flops (abs.\ / \%) & V1: 7,957 (2.96\%) out of 269,200\\
                        & V2: 8,018 (2.98\%) out of 269,200\\
BRAM tiles (abs.\ / \%) & V1: 0 (0\%) | V2: 0 (0\%)\\
DSP slices (abs.\ / \%) & V1: 0 (0\%) | V2: 0 (0\%)\\
Hidden / output accumulator width & 32-bit \\
Weight quantization & int8, scale $S_l = 127 / \max(|W_l|)$\\
Spike encoding & rate coding, seed = 42 \\
Number of timesteps $T$ & 16 \\
Latency per sample (cycles / $\mu$s) & V1: 13,647 cycles $\times$ 75~ns $\approx$ 1,024~$\mu$s \\
                                      & V2: 13,669 cycles $\times$ 6~ns  $\approx$ 82~$\mu$s\\
VHDL source modules & Top-Level: \texttt{snn\_top\_level.vhd} \\
                    & Submodules: \texttt{lif\_neuron.vhd}, \\
                    & \texttt{synapse\_accumulator.vhd}, \\
                    & \texttt{input\_spike\_driver.vhd}, \\
                    & \texttt{layer\_controller.vhd}, \\
                    & \texttt{output\_counter.vhd}, \\
                    & \texttt{argmax\_unit.vhd} \\
\bottomrule
\end{tabular}
\end{threeparttable}
}%
\end{table}

Power consumption was evaluated post-implementation using the AMD Vivado Design Suite Power Analysis tool. To establish a baseline comparable with existing FPGA accelerators in the literature, the default vectorless estimation mode is first utilized, which applies a uniform 12.5\% signal switching activity across the unconstrained design nets. However, because the vectorless estimation does not fully account for the temporal and data-driven sparsity of Spiking Neural Networks (SNNs), these baseline metrics represent a conservative upper bound in the evaluated setup.

To evaluate the estimated power characteristics of the architecture, a vector-based power analysis is also conducted. Simulated switching activity is precisely generated using the Vivado Simulator (XSim) during a behavioral simulation of the 1,000-sample MNIST test batch. The resulting toggle rates and static probabilities are recorded by exporting a Switching Activity Interchange Format (.saif) file. During the Vivado power evaluation phase, this SAIF file is applied to annotate the post-implementation design, providing high activity coverage for the internal logic and ensuring the estimation tracks the event-driven SNN operational sparsity much more accurately than static vectorless assumptions. To provide a comprehensive analysis, both vectorless and vector-based methodologies are analyzed, isolating static, dynamic, and I/O power dissipation. The Vivado report summaries for both architectural versions under both execution modes are provided in \hyperref[fig:power_reports_grid]{Appendix~A}.

\subsection{Output Classification and the Argmax Decision Core}
Following the completion of the $T = 16$ temporal integration cycles, the final classification of the MNIST digit is determined by evaluating the cumulative spike events across the 10 output-layer LIF neurons. To achieve this, an independent \texttt{output\_counter} hardware module is instantiated for each output channel to track the discrete firing events. At the end of the processing window, the central controller asserts the \texttt{calculate\_argmax} signal, halting further accumulation and activating the \texttt{argmax\_unit}. This module iterates through the 10 counter registers to identify the index corresponding to the maximum spike count. To resolve scenarios where all output neurons fail to fire, the architecture utilizes the final residual membrane potentials as a tie-breaker in the evaluated setup. This mechanism is designed to provide a prediction by utilizing the sub-threshold integration state of the LIF neurons when discrete spike outputs are absent \cite{cassidy2013}. The mathematical representation of this rate-coded winner-take-all classification strategy is expressed as:
\begin{equation}
\text{Prediction} = \begin{cases}
\underset{i \in \{0, \dots, 9\}}{\operatorname{argmax}} (S_i), & \text{if } \sum S_i > 0 \\[10pt]
\underset{i \in \{0, \dots, 9\}}{\operatorname{argmax}} (V_i[T]), & \text{otherwise}
\end{cases}
\label{eq:argmax_decision}
\end{equation}
where $S_i$ represents the total accumulated spike count for output neuron $i$, and $V_i[T]$ denotes its final membrane potential at the end of the temporal window. 

For the initial timing characterization of this baseline configuration, the hardware constraints file specifies a target clock period ($T_{\text{target}}$) of $10\text{ ns}$, corresponding to a nominal operating frequency of $100\text{ MHz}$. Following post-implementation timing analysis, the design exhibits a Worst Negative Slack (WNS) violation of $-49.380\text{ ns}$, indicating that the single-cycle execution path does not satisfy the targeted clock boundaries due to propagation delays \cite{hauck2007} in the evaluated setup. The estimated maximum operating frequency ($F_{\max}$) for this non-pipelined configuration is calculated using Equation \ref{eq:fmax_calculation}:
\begin{equation}
F_{\max} = \frac{1}{T_{\text{target}} - \text{WNS}} = \frac{1}{59.38\text{ ns}} \approx 16.8\text{ MHz}
\label{eq:fmax_calculation}
\end{equation}
Timing path analysis indicates that this delay is localized within the single-cycle combinational tree of the initial \texttt{argmax\_unit}. To execute the comparison operations across the 10 output registers and evaluate the sub-threshold membrane potentials as a tie-breaker within a single clock cycle, the synthesis engine implements more than 100 levels of look-up table (LUT) logic fabric \cite{zhang2015}. While Equation \ref{eq:fmax_calculation} identifies a theoretical minimum clock period of $59.38\text{ ns}$ ($16.8\text{ MHz}$) based on the initial timing failure, practical FPGA implementation requires a relaxed target constraint to guarantee stable routing and positive slack across all physical device corners. Consequently, the system clock constraint was conservatively relaxed to $75\text{ ns}$ ($13.33\text{ MHz}$) \cite{chu2006}. Re-synthesizing the architecture under this 75 ns constraint successfully mitigated all setup violations, yielding a stable, positive Worst Negative Slack (WNS) of $+2.02\text{ ns}$ in the evaluated setup.

Operating at this frequency, the non-pipelined VHDL SNN architecture completes a single MNIST classification pass with a per-sample hardware core runtime of $1.024\text{ ms}$ ($1024\,\mu\text{s}$). In contrast, the unoptimized sequential software reference requires approximately $84\text{ ms}$ per sample for the same workload. Consequently, despite running at a lower clock frequency than the host processor, the hardware accelerator achieves a speedup of approximately $82\times$ relative to the unoptimized software reference in the evaluated setup. However, compared to a vectorized NumPy baseline with an execution latency of approximately $277\,\mu\text{s}$ per sample on the CPU, this non-pipelined hardware configuration is outperformed by the vectorized software implementation by a factor of approximately $3.7\times$. This performance difference illustrates the limitations of a non-pipelined design and motivates the pipelined readout architecture introduced in Section 4.2. These baseline performance ratios are expressed via Equation \ref{eq:speedup_v1}:
\begin{equation}
S_{\text{V1}} = 
\begin{cases} 
    \dfrac{T_{\text{Py,unopt}}}{T_{\text{VHDL}}} \approx 82\times, & \text{vs. unoptimized} \\[10pt]
    \dfrac{T_{\text{NumPy,opt}}}{T_{\text{VHDL}}} \approx 0.27\times, & \text{vs. vectorized}
\end{cases}
\label{eq:speedup_v1}
\end{equation}
where $S_{\text{V1}}$ represents the baseline hardware speedup factor. The execution latencies per sample for the unoptimized Python pipeline ($T_{\text{Py,unopt} }=84\text{ ms/sample}$) and the optimized NumPy implementation ($T_{\text{NumPy,opt}}=277\,\mu\text{s/sample}$) are benchmarked directly against the localized VHDL accelerator core processing latency ($T_{\text{VHDL}} = 1.024\text{ ms}$).

This acceleration is driven by the spatial parallelism of the hardware. At each clock edge, all 64 hidden-layer and 10 output-layer LIF neurons concurrently execute their membrane potential updates, arithmetic leak operations, and threshold evaluations. 

To establish an empirical reference point for energy performance, the software baselines running on the Intel Core Ultra 9 processor were profiled using hardware VRM telemetry via HWInfo64~\cite{hwinfo}. Under a full inference workload, the unoptimized sequential reference exhibits a sustained average power dissipation of $56\text{ W}$ at an average core clock of $3.0\text{ GHz}$. Given its $84\text{ ms}$ per-sample execution time, it consumes $\approx 4.704\text{ Joules}$ per sample, resulting in a nominal energy throughput of $\approx 0.213\text{ samples/J}$. In comparison, the vectorized NumPy implementation utilizes the CPU's parallel execution pipelines, drawing a steady-state average power of $49\text{ W}$ at a lower core clock of $2.85\text{ GHz}$ due to the hardware's automated AVX frequency offsets \cite{intel_avx}. This software optimization represents a batched scenario where data is grouped into parallel matrix operations to maximize execution pipeline utilization. While this parallel framework minimizes execution overhead and reduces the estimated total software energy consumed to $\approx 0.0136\text{ Joules}$ per sample (yielding an estimated operational software efficiency of $\approx 73.68\text{ samples/J}$), it represents an offline throughput metric rather than a real-time edge streaming pipeline.
In the evaluated setup, post-implementation power analysis within the AMD Vivado Design Suite provides an estimated vector-based total on-chip power consumption of $0.177\text{ W}$ (with a default vectorless estimate of $0.164\text{ W}$ reported solely for structural evaluation consistency with external works).
This operational power footprint demonstrates an estimated computational energy efficiency, formalized using primary vector-based metrics in Equation \ref{eq:efficiency_v1} as:
\begin{equation}
\eta_{\text{V1}} = \frac{1}{T_{\text{VHDL}} \times P_{\text{V1}}} \approx 5.52 \times 10^3\,\text{samples/J}
\label{eq:efficiency_v1}
\end{equation}
where $\eta_{\text{V1}}$ represents the baseline hardware energy efficiency, calculated using the previously defined operational latency $T_{\text{VHDL}}$ alongside the post-implementation core power dissipation ($P_{\text{V1}} = 0.177\text{ W}$).

To evaluate the relative energy efficiency differences between the general-purpose execution pipelines and the hardware architecture \cite{Merolla2014}, the relative energy efficiency gain of the baseline hardware over both software implementations is evaluated in Equation \begin{equation}
\eta_{\text{rel}} = 
\begin{cases} 
    \dfrac{\eta_{\text{V1}}}{\eta_{\text{CPU,unopt}}} \approx 2.6 \times 10^4\times, & \text{vs. unoptimized} \\[10pt]
    \dfrac{\eta_{\text{V1}}}{\eta_{\text{CPU,opt}}} \approx 75\times, & \text{vs. vectorized}
\end{cases}
\label{eq:relative_efficiency}
\end{equation}
where $\eta_{\text{rel}}$ represents the relative hardware efficiency gain. The baseline accelerator efficiency ($\eta_{\text{V1}} = 5517$ samples/J) is compared directly against the unoptimized CPU reference ($\eta_{\text{CPU,unopt}} = 0.213$ samples/J) and the optimized, vectorized CPU baseline ($\eta_{\text{CPU,opt}} = 73.68$ samples/J), respectively. In the evaluated setup, the baseline hardware architecture indicates an estimated $\approx 26000\times$ increase in throughput-per-watt over the unoptimized reference, along with a $\approx 75\times$ efficiency gain over the vectorized software baseline running on the evaluated CPU. This demonstrates the power-scaling characteristics associated with specialized edge hardware designs, prior to introducing temporal pipelined optimizations.

Finally, physical resource utilization maps onto the target Artix-7 fabric, consuming $13\%$ of the available Look-Up Tables (LUTs) and $3\%$ of the Flip-Flop (FF) registers. Due to the multiplierless design of the architecture, weight matrices and neuron states are unrolled by the compiler into combinational logic (LUTs) and distributed registers (Flip-Flops). This mitigates memory access constraints associated with physical BRAM tiles, enabling parallel data access~\cite{neil2014}. In the evaluated setup, this configuration demonstrates that the parallel neuron math circuits and the baseline combinational argmax tree fit within the device fabric with available headroom, as observed in the post-implementation floorplan layout provided in \hyperref[sec:appendix_floorplans]{Appendix~B}. This utilization indicates physical scalability for accommodating deeper or wider network architectures on a single device.

\subsection{Microarchitectural Optimization: Pipelined Sequential Argmax (Version 2)}
To resolve the timing violation (WNS = $-49.380\text{ ns}$) identified in the baseline hardware design, a microarchitectural optimization was implemented within the \texttt{argmax\_unit}. The unrolled, purely combinational loop structures of Version 1 were replaced by a synchronous, multi-cycle Finite State Machine (FSM) control path, as detailed in the behavioral VHDL implementation of Version 2. Rather than replicating hardware comparators spatially—which increases the look-up table (LUT) logic path depth within a single clock cycle—this approach implements temporal multiplexing, re-using a single comparator slice across consecutive clock edges.

\begin{figure}
    \centering
    \resizebox{\columnwidth}{!}{%
    \begin{tikzpicture}[
        font=\sffamily\scriptsize,
        >=Stealth,
        node distance=1.2cm and 2.0cm,
        % ==================== STYLES  ====================
        state_idle/.style={
            rectangle, draw=black, fill=green!60!black, text=white, very thick, 
            minimum height=3.2em, minimum width=18em, align=center, rounded corners=4pt, 
            font=\sffamily\scriptsize
        },
        state_process/.style={
            rectangle, draw=black, fill=blue!80!black, text=white, very thick, 
            minimum height=3.2em, minimum width=18em, align=center, rounded corners=4pt, 
            font=\sffamily\scriptsize
        },
        state_eval/.style={
            rectangle, draw=black, fill=violet!70!black, text=white, very thick, 
            minimum height=3.2em, minimum width=18em, align=center, rounded corners=4pt, 
            font=\sffamily\scriptsize
        },
        state_reset/.style={
            rectangle, draw=black, fill=red!70!black, text=white, very thick, 
            minimum height=3.2em, minimum width=18em, align=center, rounded corners=4pt, 
            font=\sffamily\scriptsize
        },
        bus/.style={->, very thick, draw=black!80},
        bus_tag/.style={midway, fill=white, draw=gray!50, rounded corners=1pt, inner sep=2.5pt, font=\sffamily\tiny\bfseries, text=black, align=center}
    ]
    
    % ==================== STATE NODE PLACEMENT ====================
    \node[state_process] (compare_spike) {\textbf{COMPARE\_SPIKE\_COUNTS}};
    \node[state_process, below=1.8cm of compare_spike] (compare_membrane) {\textbf{COMPARE\_MEMBRANE\_POTENTIALS}};
    \node[state_eval, below=1.8cm of compare_membrane] (report_label) {\textbf{REPORT\_PREDICTED\_LABEL}};  
    \node[state_reset, below=1.6cm of report_label] (wait_reset) {\textbf{WAIT\_FOR\_RESET}}; 
    \node[state_idle, below=1.6cm of wait_reset] (idle) {\textbf{IDLE}};
    \node[circle, fill=black, inner sep=2.5pt] (start) at ([yshift=-1cm]idle.south) {};
    \draw[bus] (start) -- (idle.south);
    
    % ==================== VERTICAL FLOW ROUTING ====================
    \draw[bus] (compare_spike) -- node[bus_tag] {counter == 10\\AND\\Max Spike == 0\\{[Reset Counter]}} (compare_membrane);
    \draw[bus] (compare_membrane) -- node[bus_tag] {counter == 10} (report_label);
    \draw[bus] (report_label) -- node[bus_tag] {Assert prediction\_ready = '1'\\Latch predicted\_label} (wait_reset);
    \draw[bus] (wait_reset) -- node[bus_tag] {Deassert prediction\_ready = '0'} (idle);
    
    % ==================== FEEDBACK LOOPS & MARGIN ROUTING ====================
    
    % Right Margin Loop 1: Evaluate Max Spike
    \coordinate (cs_out) at ([yshift=-12pt]compare_spike.east);
    \coordinate (cs_in) at ([yshift=12pt]compare_spike.east);
    \draw[bus, rounded corners=4pt] (cs_out) -- ++(1.4,0) |- node[bus_tag, pos=0.25] {counter $<$ 10\\{[Evaluate Max Spike]}} (cs_in);
    
    % Right Margin Loop 2: Evaluate Max Potential
    \coordinate (cm_out) at ([yshift=-12pt]compare_membrane.east);
    \coordinate (cm_in) at ([yshift=12pt]compare_membrane.east);
    \draw[bus, rounded corners=4pt] (cm_out) -- ++(1.4,0) |- node[bus_tag, pos=0.25] {counter $<$ 10\\{[Evaluate Max Potential]}} (cm_in);
    
    % Right Margin Loop 3: Idle Quiescent Loop
    \coordinate (id_out) at ([yshift=-9pt]idle.east);
    \coordinate (id_in) at ([yshift=9pt]idle.east);
    \draw[bus, rounded corners=4pt] (id_out) -- ++(1.4,0) |- node[bus_tag, pos=0.25] {calculate\_argmax == '0'} (id_in);
    
    % Left Margin Loop 1: Bypass to Report Label
    \draw[bus, rounded corners=4pt] ([yshift=-6pt]compare_spike.west) -- ++(-1.8,0) |- node[bus_tag, pos=0.25] {counter == 10\\AND\\Max Spike $>$ 0} (report_label.west);
    
    % Left Margin Loop 2: Reset to Next Iteration
    \draw[bus, rounded corners=4pt] ([yshift=-6pt]idle.west) -- ++(-3.2,0) |- node[bus_tag, pos=0.25] {calculate\_argmax == '1'\\{[Reset Counters \& Labels]}} ([yshift=6pt]compare_spike.west);
    
    \end{tikzpicture}
    }%
    \caption{Finite State Machine (FSM) state diagram of the optimized sequential \texttt{argmax\_unit} (Version 2), showing multi-cycle decision branching and conditional sub-threshold membrane potential tie-breaking.}
    \label{fig:argmax_fsm}
\end{figure}
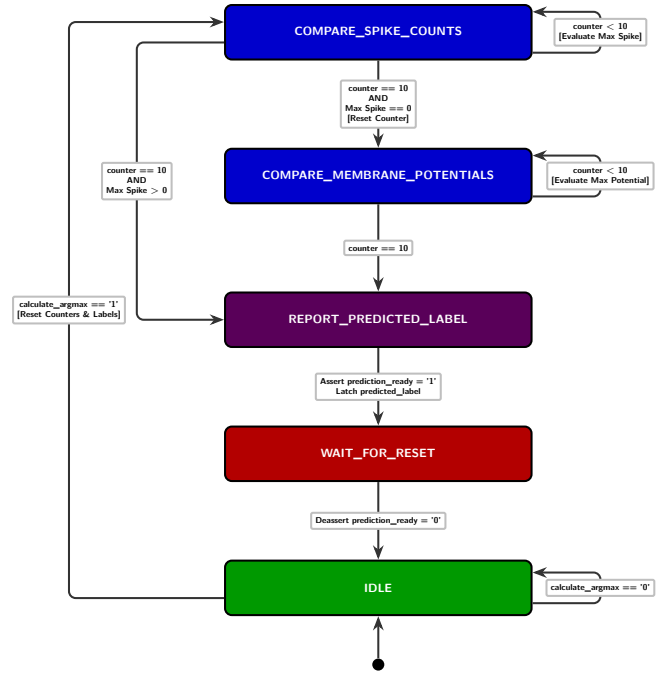

The FSM control logic is governed by five operational states: \texttt{IDLE}, \texttt{COMPARE\_SPIKE\_COUNTS}, \texttt{COMPARE\_MEMBRANE\_POTENTIALS}, \texttt{REPORT\_PREDICTED\_LABEL}, and \texttt{WAIT\_FOR\_RESET}. The corresponding state transitions and decision logic are illustrated in Fig. \ref{fig:argmax_fsm}.

In the \texttt{IDLE} state, the module holds the synchronous output flag \texttt{prediction\_ready} low while resetting the loop iteration index $i$ to $1$ and initializing the predicted label index $L_{\text{pred}}$ to $0$. The FSM awaits the assertion of the control signal \texttt{calculate\_argmax} from the central layer controller. Upon assertion, the system transitions to \texttt{COMPARE\_SPIKE\_COUNTS}, where the FSM iterates through the 10 output-channel spike accumulation registers over a 9-cycle window. At each rising clock edge, the circuit evaluates the conditional inequality defined in \eqref{eq:sequential_spike_compare}:
\begin{equation}
L_{\text{pred}} \leftarrow 
\begin{cases} 
    i, & \text{if } S_i > S_{L_{\text{pred}}} \\
    L_{\text{pred}}, & \text{otherwise}
\end{cases}
\label{eq:sequential_spike_compare}
\end{equation}
where $S_i$ denotes the accumulated spike count for output channel $i$.

Once $i$ reaches $10$, the FSM executes a conditional control branch. If the maximum accumulated spike count is strictly greater than zero ($\max(S_i) > 0$), the tie-breaker mechanism is bypassed, and the FSM transitions directly to \texttt{REPORT\_PREDICTED\_LABEL}. Conversely, if no spikes were recorded across the output layer, the FSM branches to \texttt{COMPARE\_MEMBRANE\_POTENTIALS} to resolve the inference tie-condition. In this state, the FSM re-initializes the iteration index $i$ to $1$ and executes a secondary comparison pass over 9 clock cycles, evaluating sub-threshold integration states as defined in \eqref{eq:sequential_potential_compare}:

\begin{equation}
L_{\text{pred}} \leftarrow 
\begin{cases} 
    i, & \text{if } V_i[T] > V_{L_{\text{pred}}}[T] \\
    L_{\text{pred}}, & \text{otherwise}
\end{cases}
\label{eq:sequential_potential_compare}
\end{equation}

where $V_i[T]$ denotes the final membrane potential of output channel $i$ at the terminal time step $T$. 

Following the completion of the tie-breaking pass, the FSM enters \texttt{REPORT\_PREDICTED\_LABEL}, latching the computed $L_{\text{pred}}$ onto the primary output bus and asserting \texttt{prediction\_ready} high. The sequence concludes with a single-cycle \texttt{WAIT\_FOR\_RESET} state, ensuring a synchronized handshake before returning to the \texttt{IDLE} state. By distributing execution latency over a variable window of $12$ to $22$ clock cycles, this microarchitectural change isolates the critical path within a single comparator block, facilitating significantly improved timing closure.

\subsection{Post-Optimization Implementation Results}

\begin{table*}
\caption{Cross-Platform Hardware Performance and Efficiency Summary}
\label{tab:hardware_performance_summary}
\centering
\resizebox{\textwidth}{!}{%
\footnotesize
\setlength{\tabcolsep}{12pt}
\begin{tabular}{lcccc}
\toprule
\textbf{Metric / Parameter} & 
\textbf{\makecell{CPU Baseline\\(Unopt. Python)}} & 
\textbf{\makecell{CPU Baseline\\(Vectorized NumPy)}} & 
\textbf{\makecell{Hardware V1\\(VHDL Baseline)}} & 
\textbf{\makecell{Hardware V2\\(Pipelined VHDL)}} \\ 
\midrule
Target Device Platform & Intel Core Ultra 9 275HX & Intel Core Ultra 9 275HX & Artix-7 FPGA & Artix-7 FPGA \\
Operating Clock Frequency, $f_{\text{CLK}}$ & $3.0\text{ GHz}$ & $2.85\text{ GHz}$ & $13.3\text{ MHz}$ & $167\text{ MHz}$ \\
Target Clock Period, $T_{\text{target}}$ & — & — & $75.000\text{ ns}$ & $6.000\text{ ns}$ \\
Worst Negative Slack (WNS) & — & — & $+2.023\text{ ns}$ & $+0.090\text{ ns}$ \\ 
\midrule
Latency per Sample, $T$ & $84.00\text{ ms}$ & $277.00\,\mu\text{s}$ & $1.024\text{ ms}$ & $82.00\,\mu\text{s}$ \\
Speedup Factor, $S_{\text{rel}}$ (vs. Unopt. CPU) & $1.0\times$ & $3.0 \times 10^2\times$ & $82.0\times$ & $1.02 \times 10^3\times$ \\
Speedup Factor, $S_{\text{rel}}$ (vs. Vectorized CPU) & $0.003\times$ & $1.0\times$ & $0.27\times$ & $3.4\times$ \\
Speedup Factor, $S_{\text{rel}}$ (vs. Hardware V1) & — & — & $1.0\times$ & $12.5\times$ \\ 
\midrule
Core Power Dissipation, $P$ & $56.00\text{ W}$ & $49.00\text{ W}$ & $0.177\text{ W}$ & $0.336\text{ W}$ \\
Estimated Batch Energy, $E_{\text{batch}}$ ($10^3$ samples) & $4700.0\text{ J}$ & $13.6\text{ J}$ & $0.18\text{ J}$ & $0.028\text{ J}$ \\
Estimated Energy Efficiency, $\eta$ & $0.213\text{ samples/J}$ & $73.68\text{ samples/J}$ & $5.52 \times 10^3\text{ smpl/J}$ & $3.63 \times 10^4\text{ smpl/J}$ \\
Efficiency Gain, $\eta_{\text{gain}}$ (vs. Unopt. CPU) & $1.0\times$ & $3.46 \times 10^2\times$ & $2.60 \times 10^4\times$ & $1.70 \times 10^5\times$ \\
Efficiency Gain, $\eta_{\text{gain}}$ (vs. Vectorized CPU) & $0.003\times$ & $1.0\times$ & $75.0\times$ & $493.0\times$ \\
Microarchitectural Efficiency Gain ($\eta_{\text{V2}}/\eta_{\text{V1}}$) & — & — & $1.0\times$ & $6.6\times$ \\ 
\bottomrule
\end{tabular}
}%
\end{table*}

The transition to a multi-cycle Finite State Machine (FSM) addressed the combinational bottleneck of the argmax unit. Synthesizing the revised architecture (Version 2) on the Artix-7 FPGA with a target clock period of 6~ns yields a positive Worst Negative Slack (WNS) of +0.09~ns. This indicates that the design operates at a maximum frequency ($F_{\text{max}}$) of approximately 167~MHz in the evaluated setup. At this clock rate, the cycle-accurate simulated latency to process a single test sample is evaluated at $82\,\mu\text{s}$. To evaluate this optimization across both software execution paradigms, the performance acceleration factors relative to the 5.4~GHz CPU are presented in Equation \ref{eq:speedup_v2}:
\begin{equation}
S_{\text{V2}} = 
\begin{cases} 
    \dfrac{T_{\text{Py,unopt}}}{T_{\text{VHDL,V2}}} \approx 1.02 \times 10^3\times, & \text{vs. unoptimized} \\[10pt]
    \dfrac{T_{\text{NumPy,opt}}}{T_{\text{VHDL,V2}}} \approx 3.4\times, & \text{vs. vectorized}
\end{cases}
\label{eq:speedup_v2}
\end{equation}
where $S_{\text{V2}}$ denotes the speedup performance achieved by the optimized, multi-cycle pipelined architecture variant. Execution latencies are benchmarked directly against the accelerated VHDL core runtime ($T_{\text{VHDL,V2}} = 82\,\mu\text{s}$). This represents an approximate $1000\times$ speedup over the unoptimized sequential reference and a $3.4\times$ speedup over the vectorized NumPy baseline. With an execution latency of $82\,\mu\text{s}$ per sample, the 167~MHz hardware core demonstrates a performance advantage over the CPU running vectorized code in the evaluated setup, while simultaneously achieving a $12.5\times$ speedup over the initial single-cycle hardware layout (Version 1). This comparison accounts for the fact that the software baseline relies on the accumulation of thousands of samples into memory to execute simultaneously via parallel SIMD matrix operations. In edge deployments—such as live sensor processing or streaming video—such batching may not be feasible, requiring the CPU to process inputs sequentially where interpreter and thread context overhead can increase latency. In contrast, the proposed VHDL design operates as a sequential streaming architecture, processing samples individually with a cycle-accurate simulated latency of $82\,\mu\text{s}$ per image, outperforming the batch throughput of the evaluated CPU setup.

To empirically validate this streaming constraint on the host platform, an additional software experiment was conducted where the vectorized NumPy core was forced to process the test samples sequentially via an outer loop, thereby disabling the look-ahead batching mechanism. Under this streaming configuration, the evaluated CPU's execution time is estimated at $375\,\mu\text{s}$ per sample while drawing approximately $51\text{ W}$ of continuous steady-state power. Because the sequential overhead limits the utilization of the CPU's wide SIMD registers between iterations, the total computational energy consumed is estimated at $\approx 0.0191\text{ Joules}$ per sample ($\approx 52.3\text{ samples/J}$). When compared against this non-batched streaming baseline, the custom 167~MHz FPGA accelerator demonstrates an approximate $4.6\times$ architectural speedup, while achieving a $694\times$ energy efficiency gain in the evaluated setup. This indicates that when general-purpose architectures are forced to operate under localized edge constraints without offline batching, the spatial computing layout of the specialized SNN hardware offers an operational advantage.

Operating at a higher frequency naturally incurs higher dynamic power consumption, estimated at a SAIF-based estimated total on-chip power of $0.336\text{ W}$ for Version 2 (with a baseline vectorless estimate of $0.637\text{ W}$ reported solely for structural evaluation consistency with external works). However, the reduction in execution time improves the overall energy metrics of the system. The energy performance of this multi-cycle optimized core is evaluated using the vector-based metrics as:
\begin{equation}
\eta_{\text{V2}} = \frac{1}{T_{\text{VHDL,V2}} \times P_{\text{V2}}} \approx 3.63 \times 10^4\,\text{samples/J}
\label{eq:efficiency_v2}
\end{equation}
where $\eta_{\text{V2}}$ represents the energy efficiency of the optimized, multi-cycle pipelined accelerator core. This performance metric is derived using the previously defined operational latency $T_{\text{VHDL,V2}}$ alongside the corresponding post-implementation total power dissipation ($P_{\text{V2}} = 0.336\text{ W}$), yielding a peak operational efficiency of approximately $36,300\text{ samples/J}$.

To evaluate both the internal microarchitectural improvements and the cross-domain aspects of shifting to custom neuromorphic spatial fabrics, the relative energy efficiency gains are presented in Equation \ref{eq:efficiency_ratio}:
\begin{equation}
\eta_{\text{gain}} = 
\begin{cases} 
    \dfrac{\eta_{\text{V2}}}{\eta_{\text{V1}}} \approx 6.6\times, & \text{vs. V1 baseline} \\[10pt]
    \dfrac{\eta_{\text{V2}}}{\eta_{\text{CPU,opt}}} \approx 493\times, & \text{vs. vectorized CPU}
\end{cases}
\label{eq:efficiency_ratio}
\end{equation}
where $\eta_{\text{gain}}$ represents the relative energy efficiency scaling factor achieved by the pipelined execution model.
The optimized multi-cycle design is therefore estimated to operate more than six times as efficiently as its unpipelined hardware predecessor in the evaluated setup. Furthermore, when compared against the optimized parallel software baseline, the custom VHDL core demonstrates an approximate $493\times$ aggregate energy-efficiency gain, indicating that specialized edge neuromorphic accelerators can deliver the required inference throughput while operating below the sustained high-power boundaries ($49\text{ W} - 56\text{ W}$) of the evaluated general-purpose von Neumann computing architecture.

The design also maintains a low resource footprint across the AMD Artix-7 fabric \cite{Lopez-Asunci2024}. Post-implementation utilization metrics reveal that the SNN architecture consumes 12\% of the available Look-Up Tables (LUTs) and 3\% of the Flip-Flops (FFs) on the Artix-7 device. This is lower than the previous design's 13\% LUT utilization, indicating that pipelining the argmax tree reduced routing congestion while improving both clock speed and spatial efficiency. Furthermore, this optimization preserves the 0\% utilization of physical block RAM (BRAM) tiles observed in the baseline design. All quantized weights and dynamic neuron states remain contained within the localized memory buffers mapped to the distributed logic fabric. This indicates that the transition to a sequential, multi-cycle argmax control path yields its timing closure and spatial advantages without introducing additional resource overheads or dual-port memory access bottlenecks. This level of utilization demonstrates the spatial efficiency of the multi-cycle redesign, providing available resources for larger network scales on the same device, as detailed in the spatial floorplan views in \hyperref[sec:appendix_floorplans]{Appendix~B}.

To explicitly isolate the effects of the proposed microarchitectural changes, Table \ref{tab:hardware_performance_summary} serves as an ablation comparison between the baseline single-cycle readout (Version 1) and the multi-cycle pipelined FSM readout (Version 2). The replacement of the wide combinational argmax tree with a temporal, multi-cycle FSM is the sole structural difference between these two hardware iterations. This single modification directly alleviated the system's critical path bottleneck, enabling the maximum clock frequency to scale from 13.3 MHz to 167 MHz and thereby reducing per-sample execution latency from 1.024 ms to 82 $\mu$s. Furthermore, while operating at a higher frequency naturally increased the SAIF-based estimated on-chip power from 0.177 W to 0.336 W, the drastic reduction in execution time ultimately yielded a $6.6\times$ improvement in estimated energy efficiency per sample. Spatially, transitioning from a fully unrolled comparator tree to a sequential, single-comparator FSM reduced overall LUT utilization from 12.72\% to 11.95\% by eliminating significant interconnect routing congestion.

\subsection{Comparison with State-of-the-Art FPGA SNN Accelerators}

\begin{table*}
\centering
\caption{Comparison of MNIST SNN Inference Accelerators on Programmable Logic}
\label{tab:comparison}
\resizebox{\textwidth}{!}{%
\begin{tabular}{p{3.5cm}cccc}
\hline
\textbf{Metric} & \makecell[c]{\textbf{Proposed Work} \\ \textbf{(Version 2)}} & \makecell[c]{\textbf{Li et al.} \\ \textbf{(2021)}} & \makecell[c]{\textbf{Carpegna et al.} \\ \textbf{(2024)}} & \makecell[c]{\textbf{Harmeling et al.} \\ \textbf{(2026)}} \\ \hline
FPGA Process Node & 28 nm & 28 nm & 28 nm & 28 nm \\
Device Family & Artix-7 & Virtex-7 & Zynq-7000 & Artix-7 \\
Network Topology & $784 - 64 - 10$ & $784 - 200 - 100 - 10$ & $784 - 128 - 10$ & $784 - 100 - 10$ \\
Neuron Model     & LIF               & LIF               & LIF               & SRC \\
Timestep Count ($T$) & 16 & $-$ & 100 & 220 \\
Input Encoding & Rate Coding & Rate Coding & Poisson Rate & Bernoulli Rate \\
Test Set Evaluated & Full (10k) & Full (10k) & Full (10k) & Full (10k) \\
Clock Frequency  & $167\,\text{MHz}$ & $100\,\text{MHz}$ & $100\,\text{MHz}$ & $100\,\text{MHz}$ \\
Accuracy (MNIST) & $96.70\%$& $92.93\%$         & $93.85\%$         & $96.31\%$ \\
Latency per Image& $82\,\mu\text{s}$ & $3.15\,\text{ms}$ & $780\,\mu\text{s}$ & $1.748\,\text{ms}$ \\
Energy per Image \\(Vectorless)& $0.053\,\text{mJ}$& $5.04\,\text{mJ}$ & $0.14\,\text{mJ}$ & $1.972\,\text{mJ}$ \\
 Energy per Image \\(Vector\_based SAIF)& $0.028\,\text{mJ}$& $-$& $-$&$-$\\
\hline
\end{tabular}
}%
\end{table*}

When evaluated against contemporary Leaky Integrate-and-Fire (LIF) implementations, the proposed architecture demonstrates a trade-off in processing latency and estimated energy efficiency. Table \ref{tab:comparison} summarizes the performance metrics of our proposed architecture alongside existing implementations, highlighting the differences in implementation platforms and evaluation methodologies. It should be noted that target platforms, network topologies, datasets, and power-measurement methodologies differ across the evaluated literature. Consequently, the performance comparison presented in this section is indicative within the evaluated setup rather than strictly controlled.

The 2024 \textit{Spiker+} framework by Carpegna et al.~\cite{carpegna2024spiker} introduced a library of ``lean'' neuron blocks designed for modern FPGA and System-on-Chip (SoC) architectures. Operating at 100 MHz on an AMD Xilinx Zynq-7020 device (a 28~nm process node), this framework achieves a classification accuracy of 93.85\% with an inference latency of 780 µs and an estimated energy consumption of 0.14 mJ per image (vectorless).  

Comparisons can also be made with architectures featuring alternative hardware scheduling and learning paradigms. The 2021 adaptive clock/event-driven processor with on-chip STDP capability proposed by Li et al.~\cite{li2021fast} utilizes a multi-layer topology operating at 100 MHz on a Virtex-7 FPGA (a 28~nm process node). Their implementation reports an inference latency of 3.15 ms, an estimated vectorless energy profile of 5.04 mJ per image, and an MNIST classification accuracy of 92.93\%. 

The proposed multi-cycle architecture (Version 2) utilizes a synchronous execution path designed for feedforward inference. Operating at a clock frequency of 167 MHz on a 28~nm Artix-7 device, it processes a single digit over 16 timesteps in a cycle-accurate simulated latency of 82~µs. This represents an approximate 9.5$\times$ reduction in latency compared to the \textit{Spiker+} framework and a 38$\times$ reduction compared to the model by Li et al. By mapping synaptic weights into localized internal memory buffers and using a hidden layer of 64 neurons, the design minimizes routing complexity. The resulting energy consumption per inference is estimated at 0.053~mJ using vectorless metrics and 0.028~mJ using vector-based (SAIF) telemetry. Furthermore, the design achieves a classification accuracy of 96.70\%, compared to the 92.93\% reported by Li et al.

The proposed design can also be evaluated against implementations that employ biologically rich models. A 2026 implementation by Harmeling et al.~\cite{harmeling2026src} evaluated Spiking Recurrent Cells (SRC) on an Artix-7 FPGA. Their model utilizes recurrent loops and multi-step dynamics to achieve continuous temporal spike profiles. Their reported cell configuration achieves an MNIST classification accuracy of $96.31\%$, with an inference latency of $1.748\,\text{ms}$ and an estimated $1.972\,\text{mJ}$ of vectorless energy per digit.

The proposed core demonstrates an approximate $21\times$ reduction in inference latency compared to this recurrent architecture, while achieving a $0.39\%$ higher classification accuracy and a lower estimated vectorless energy consumption. Implementing the reduction operation as a synchronous, pipelined sequential argmax Finite State Machine (FSM) instead of a wide, combinational reduction tree reduces routing complexity. By breaking the combinational delay paths into discrete timing boundaries, the architecture enables an operating frequency of $167\,\text{MHz}$ on the Artix-7 fabric, improving throughput and energy metrics compared to the alternative hardware implementations.

\section{Conclusion}
\label{sec:conclusion}

\subsection{Architectural Extensions and Future Improvements}

\subsubsection{Frequency Scaling and Arithmetic Pipelining}
While the current multi-cycle optimization achieves timing closure at a target clock period of $6\text{ ns}$ ($F_{\text{max}} \approx 167\text{ MHz}$), increasing the operating frequency above $200\text{ MHz}$ requires reducing the clock period to $5\text{ ns}$ or below. Physical synthesis and timing analysis indicate that at this frequency threshold, the primary critical path shifts from the classification logic to the combinational arithmetic within the Leaky Integrate-and-Fire module. Specifically, the back-to-back calculation of the leak decay and synaptic current integration—defined by the integer expression in Equation \ref{eq:lif_neuron}—increases propagation delay within a single clock cycle. To address this timing path, future implementations can incorporate microarchitectural pipelining within the neuron's core. By inserting intermediate synchronous registers to decouple the arithmetic right-shift subtraction from the subsequent 32-bit synaptic current accumulation, the combinational path delay can be reduced, enabling operation above $200\text{ MHz}$.

\subsubsection{Preempting the Secondary Bottleneck: Dual-Track Parallel Argmax}
Once the arithmetic path of the LIF neurons is pipelined, synthesis indicates that the sequential readout may become the next timing bottleneck. The current Version 2 design utilizes a two-stage sequential control approach: it evaluates the 10 output channels over a 10-cycle window to determine spike counts and conditionally initiates an additional 10-cycle loop to evaluate sub-threshold membrane potentials if no spikes are recorded. This approach results in an execution latency of up to 22 clock cycles. 

To bound the execution latency to a fixed 10-cycle window, a \textit{Dual-Track Parallel Argmax Engine} is introduced. Instead of executing the two evaluation sweeps sequentially, this architecture processes both metrics concurrently. In each clock cycle $i$, the engine compares the incoming channel spike counts ($S_i$ vs. $S_{L_{\text{spk}}}$) and the terminal membrane potentials ($V_i[T]$ vs. $V_{L_{\text{pot}}}[T]$). Two independent tracking registers evaluate these paths across the 10-cycle sweep to update the running spike label index ($L_{\text{spk}}$) and potential label index ($L_{\text{pot}}$), as formalized in Equations \ref{eq:dual_track_spike} and \ref{eq:dual_track_pot}:

\begin{equation}
L_{\text{spk}} \leftarrow 
\begin{cases} 
    i, & \text{if } S_i > S_{L_{\text{spk}}} \\
    L_{\text{spk}}, & \text{otherwise}
\end{cases}
\label{eq:dual_track_spike}
\end{equation}

\begin{equation}
L_{\text{pot}} \leftarrow 
\begin{cases} 
    i, & \text{if } V_i[T] > V_{L_{\text{pot}}}[T] \\
    L_{\text{pot}}, & \text{otherwise}
\end{cases}
\label{eq:dual_track_pot}
\end{equation}

At the end of the 10th clock cycle, a global zero-spike detection flag evaluates the total layer activity $\sum S_k$ to select the final predicted label index ($L_{\text{pred}}$) via a hardware multiplexer structure:

\begin{equation}
L_{\text{pred}} \leftarrow 
\begin{cases} 
    L_{\text{spk}}, & \text{if } \sum S_k > 0 \\[6pt]
    L_{\text{pot}}, & \text{if } \sum S_k = 0
\end{cases}
\label{eq:dual_track_mux}
\end{equation}

This approach introduces a microarchitectural trade-off: evaluating two parallel inequalities concurrently increases the combinational workload and the physical Look-Up Table (LUT) logic depth within the argmax module. However, when combined with the proposed neuron arithmetic pipelining, this localized increase in logic depth does not affect the global critical path. This dual-track strategy eliminates the conditional 10-cycle overhead, ensuring a deterministic classification window.

\subsubsection{Ultra-High-Frequency Scaling: Synchronous Logarithmic Binary Tree Reduction}
For higher operating frequencies where a sequential comparator restricts the master clock edge, the execution window of the argmax unit can be compressed from 10 clock cycles down to 4 cycles. This is achieved by restructuring the 10-element sweep into a \textit{Synchronous Logarithmic Binary Tree Reduction} network. 

Instead of evaluating channels sequentially, the 10 channels are grouped into five pairs during the first clock cycle. Parallel comparators evaluate these pairs simultaneously, and on the subsequent clock edge, the five intermediate winners are registered into pipeline registers. These winners are then paired again to recursively reduce the pool until the global maximum index is determined.

Substituting the $O(N)$ linear complexity with an $O(\log_2 N)$ binary reduction tree minimizes the processing latency of the classification stage. This architectural shift introduces a hardware trade-off, increasing resource utilization and routing density on the FPGA fabric due to the concurrent instantiation of parallel comparators. However, by restructuring an unrolled spatial tree into a synchronously registered pipeline, the critical path is bounded to a single comparator stage. This ensures the classification engine can keep pace with the neuron processing pipelines, enabling higher operating frequencies for neuromorphic edge accelerators.

\subsection{Limitations and Boundary Conditions}
\label{sec:limitations}
The current limitations of this work are summarized below to define the scope of the evaluation and guide future work:
\begin{itemize}
    \item \textbf{Dataset scope:} The evaluation is currently limited to the standard MNIST dataset. Extending the microarchitectural validation to more complex or natively neuromorphic benchmarks (e.g., Fashion-MNIST, N-MNIST, and DVS-Gesture) is a key objective.
    \item \textbf{Inference-only:} The architecture supports forward inference exclusively; on-chip or online learning is not supported.
    \item \textbf{Simulation-based power estimation:} While the power and energy metrics utilize post-implementation vector-based SAIF profiles generated from behavioral simulation, these represent CAD-estimated values rather than direct physical board-level silicon measurements.
    \item \textbf{Scalability:} Time-multiplexed execution scales linearly with input size and timesteps, which may limit inference throughput for significantly larger models.
    \item \textbf{Coding scheme:} The rate coding mechanism can be inefficient for sparse, event-based workloads where temporal or rank-order coding would better exploit data sparsity.
    \item \textbf{Deployment pipeline:} Data and weights are loaded at compile time via VHDL impure functions; a complete runtime deployment pipeline with external memory or host streaming interfaces is not yet implemented.
\end{itemize}

\subsection{Concluding Remarks}
This work presents the design, optimization, and FPGA implementation of an inference-only Spiking Neural Network (SNN) for MNIST classification on an AMD Artix-7 FPGA platform. By utilizing an integer-driven Leaky Integrate-and-Fire (LIF) neuron model with 32-bit registers, the architecture avoids overflow hazards and matches the software reference model without requiring neuron or synapse pruning. To mitigate physical routing constraints and combinational delays, a centralized Finite State Machine (FSM) controller manages a serialized 1-bit input spike distribution network mapped to local memory buffers.

The principal novelty of this work is the holistic integration of these subsystems into a cohesive neuromorphic pipeline. As a key enabler of this integrated architecture, the transition from a single-cycle combinational argmax tree (Version 1) to a multi-cycle sequential argmax unit (Version 2) eliminates the dominant readout bottleneck. This optimization achieves timing closure, improving the Worst Negative Slack (WNS) from $-49.380\text{ ns}$ to $+0.090\text{ ns}$ at a maximum frequency ($F_{\text{max}}$) of $167\text{ MHz}$. At this frequency, the core processes a single sample in $82\,\mu\text{s}$ ($0.082\text{ s}$ per 1,000-sample batch), representing an approximately $1000\times$ speedup over the sequential Python reference and a $\approx 3.4\times$ throughput improvement over a parallelized NumPy baseline. While the software baseline requires batch processing to maximize hardware utilization, the custom VHDL core operates as a real-time streaming architecture, processing inputs sample-by-sample with a latency of $\approx 82\,\mu\text{s}$ per image; however, it is important to note that this execution time scales linearly with input size and the number of operational timesteps.

Although operating at a higher frequency increases the total on-chip power to $0.336\text{ W}$ (compared to a baseline vectorless estimate of $0.637\text{ W}$), the reduced latency improves estimated energy efficiency by over $6.6\times$ compared to Version 1, achieving an estimated $\approx 36300\,\text{samples/J}$. Compared to the parallelized NumPy execution on a CPU drawing $49\text{ W}$, the hardware architecture achieves a $\approx 493\times$ energy-efficiency gain, demonstrating the efficiency of specialized edge neuromorphic accelerators over conventional von Neumann computing systems. The design is suitable for real-world streaming applications where data batching is impractical. Combined with a low resource footprint consuming $12\%$ of available LUTs and $3\%$ of Flip-Flops, the core maintains a classification accuracy of $96.70\% \pm 0.04\%$ across five random seeds on the 10,000-sample MNIST test set.

Overall, this work demonstrates that low-cost, spatial neuromorphic hardware can deliver high computational throughput per Watt within embedded streaming constraints. This efficiency is rigorously validated by the Vivado post-implementation power estimates and post-implementation timing analyses detailed in Section \ref{sec:results}. By mitigating routing congestion via time-multiplexed spike broadcasting and eliminating the readout critical path through multi-cycle pipelining, the proposed microarchitecture establishes a spatially efficient paradigm for real-time edge intelligence on resource-constrained programmable fabrics, carefully balancing parallel throughput with the latency trade-offs inherent to time-multiplexing.

\section*{Code and Data Availability}
To support experimental reproducibility and facilitate transparent verification of the reported hardware metrics, the complete implementation framework developed in this study is publicly accessible at \url{https://github.com/RezaAnsari471/SNN-TimeMultiplexed-Pipelined-Accelerator}.

\section*{Appendix A: Comprehensive Power Analysis Reports}
\label{sec:appendix_power}
This appendix presents the complete post-implementation power analysis reports generated via the AMD Vivado Design Suite. To accurately capture the sparse, event-driven nature of the proposed Spiking Neural Network, power dissipation is evaluated using both default vectorless estimation and vector-based profiling driven by a Switching Activity Interchange Format (.saif) simulation. 

\begin{figure}[!htbp]
    \centering
    % Top Row: V1
    \begin{subfigure}[b]{0.35\columnwidth}
        \centering
        \includegraphics[width=\columnwidth]{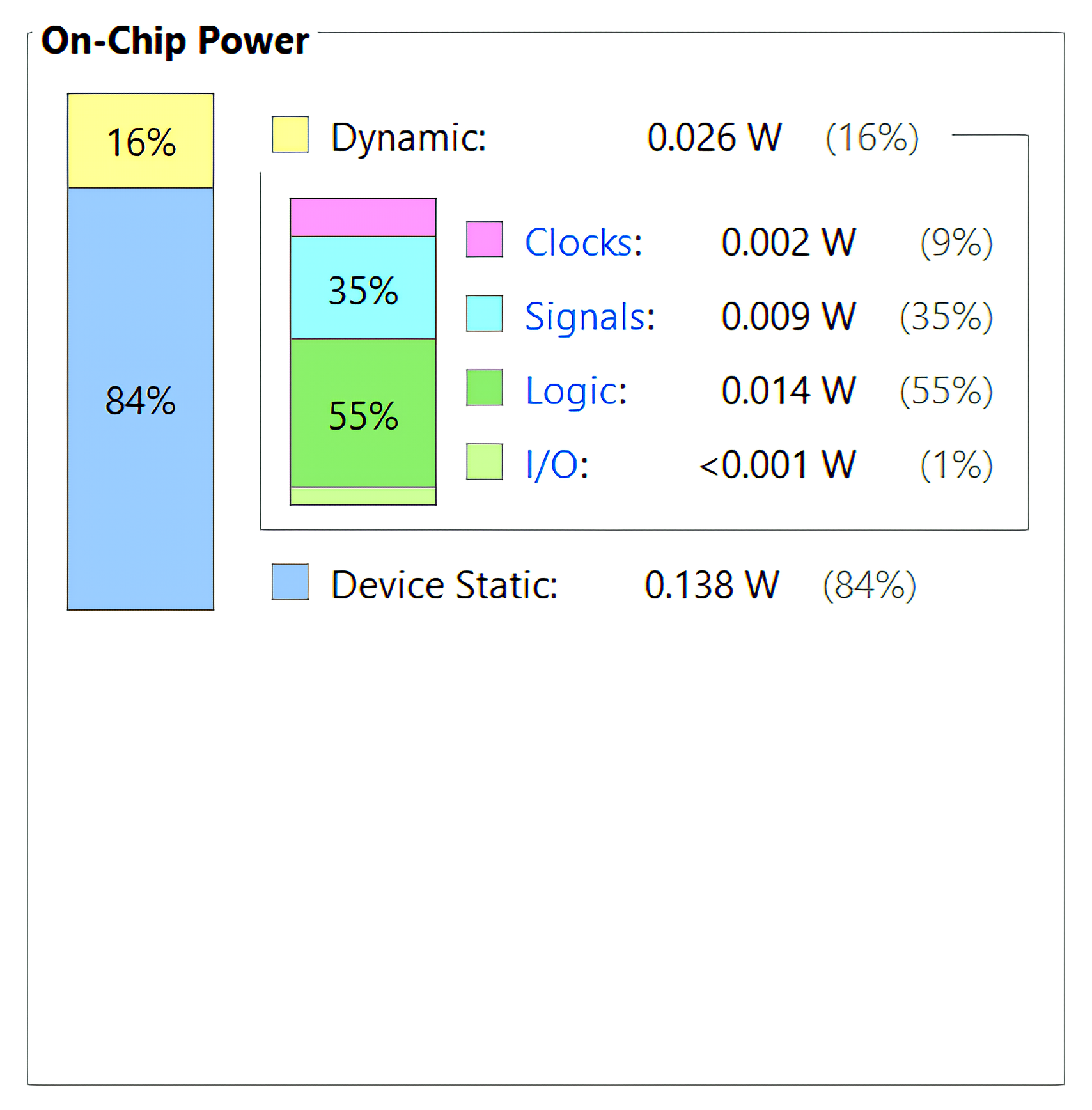}
        \caption{V1: Vectorless Estimation}
        \label{fig:v1_vectorless}
    \end{subfigure}
    \quad
    \begin{subfigure}[b]{0.35\columnwidth}
        \centering
        \includegraphics[width=\columnwidth]{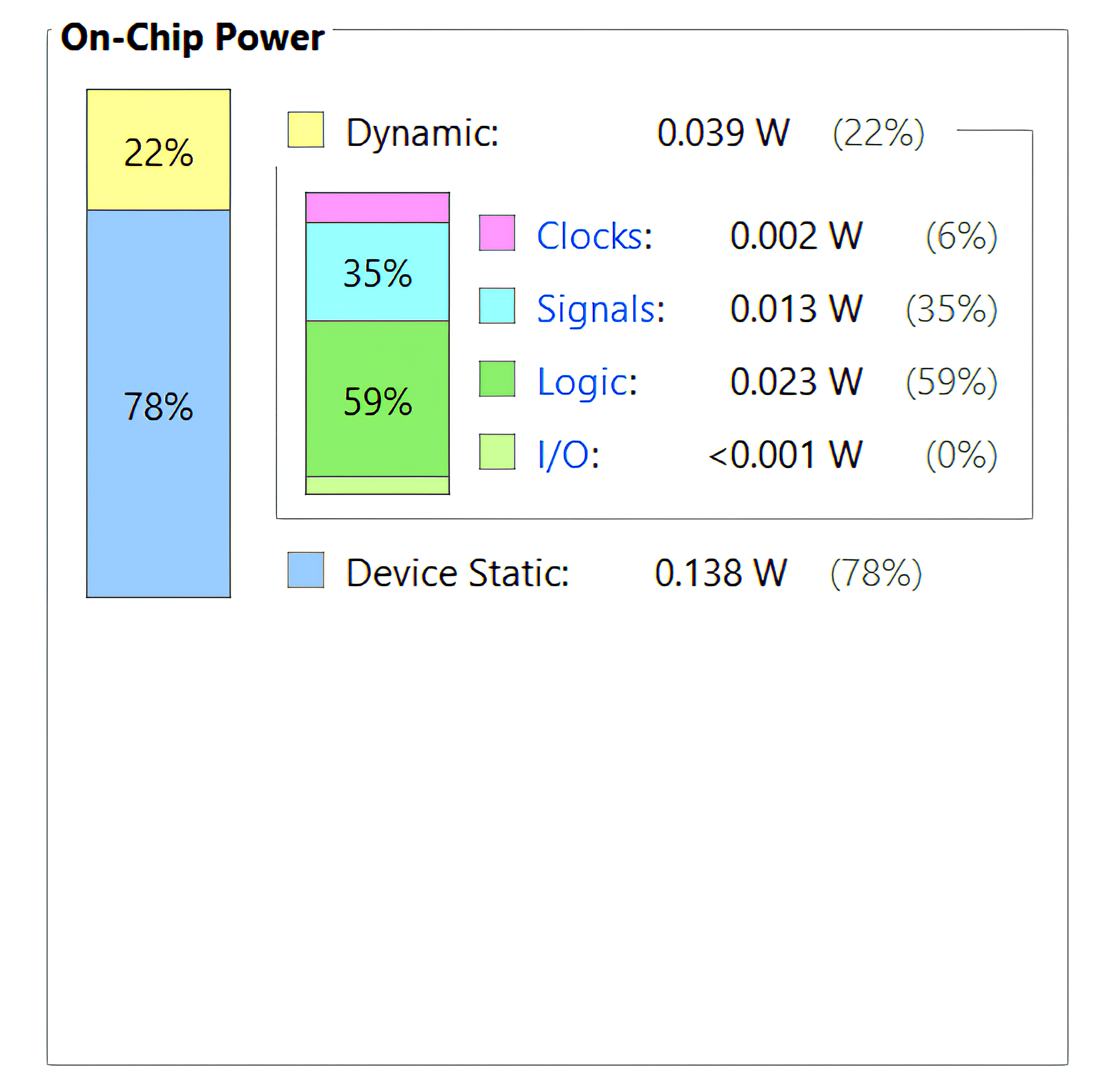}
        \caption{V1: Vector-Based (SAIF) Profiling}
        \label{fig:v1_vectorbased}
    \end{subfigure}
    
    \vspace{1em}
    
    % Bottom Row: V2
    \begin{subfigure}[b]{0.35\columnwidth}
        \centering
        \includegraphics[width=\columnwidth]{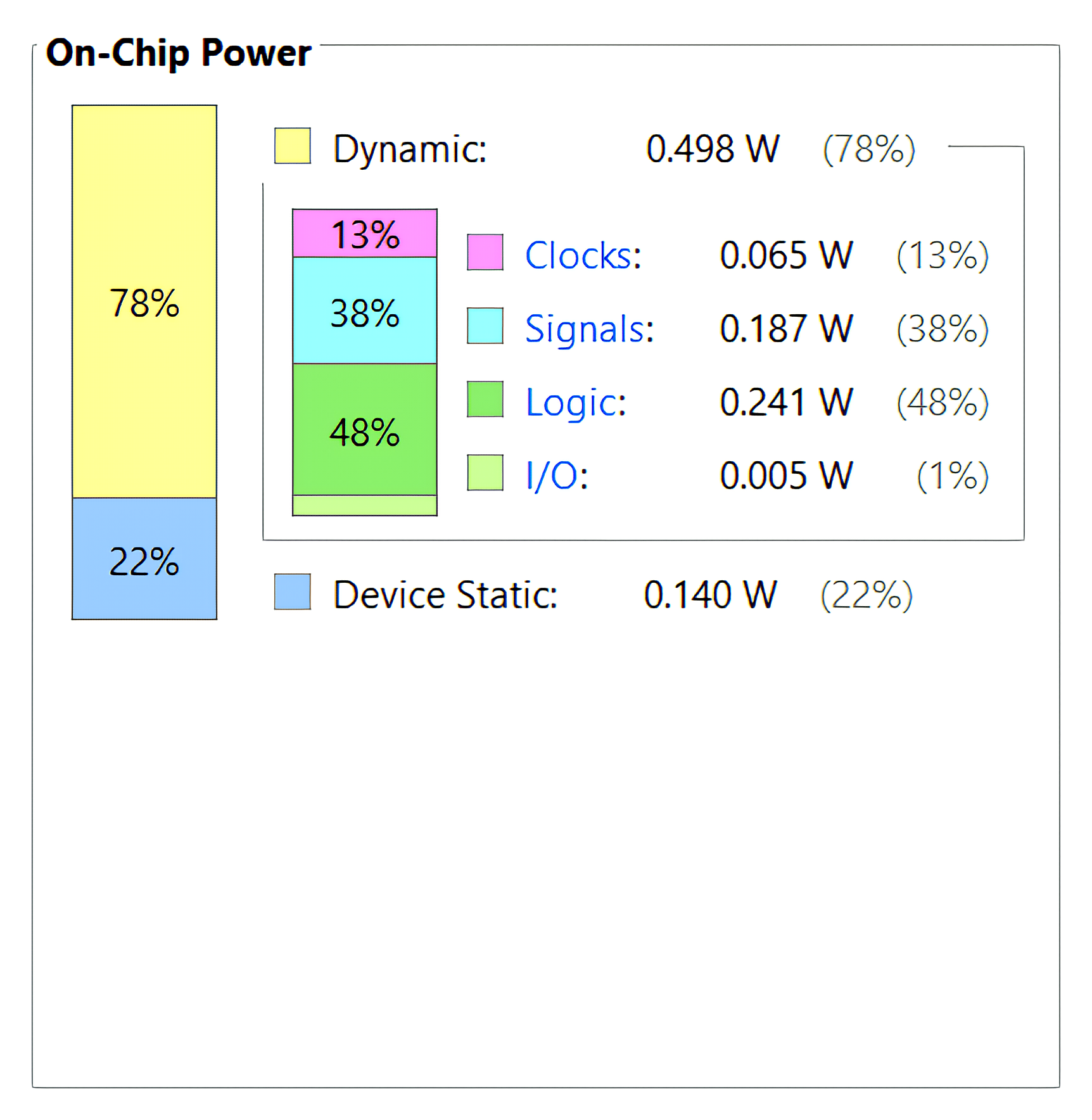}
        \caption{V2: Vectorless Estimation}
        \label{fig:v2_vectorless}
    \end{subfigure}
    \quad
    \begin{subfigure}[b]{0.35\columnwidth}
        \centering
        \includegraphics[width=\columnwidth]{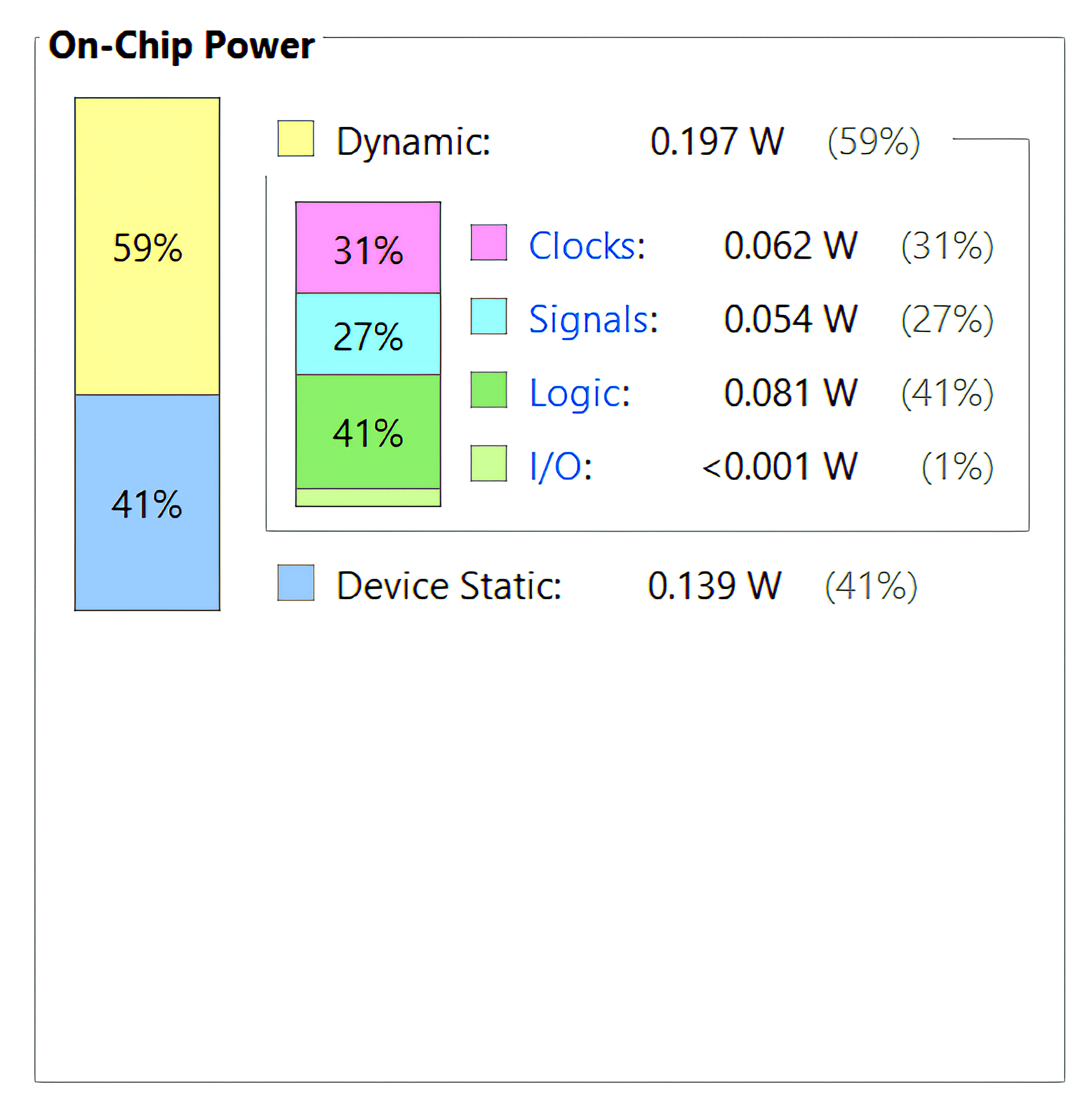}
        \caption{V2: Vector-Based (SAIF) Profiling}
        \label{fig:v2_vectorbased}
    \end{subfigure}
    \caption{Post-implementation power analysis results reported by Vivado.}
    \label{fig:power_reports_grid}
\end{figure}

The visual power distribution summaries are provided in Figure \ref{fig:power_reports_grid}. The top row depicts the analysis for the baseline non-pipelined architecture (V1), while the bottom row illustrates the optimized multi-cycle pipelined architecture (V2). Each row contrasts the baseline vectorless estimation against the vector-based behavioral profiling. 

A detailed numerical breakdown of these operating conditions is summarized in Table \ref{tab:detailed_power_breakdown}. The data highlights the critical importance of vector-based analysis when evaluating neuromorphic hardware on FPGAs. For instance, in the optimized V2 architecture operating at 167 MHz, the default vectorless estimation predicts a total on-chip power of 0.637 W, driven largely by an overestimated dynamic logic power dissipation of 0.241 W. However, when actual event-driven switching activity is applied via the SAIF profile, the total power drops by nearly half to 0.336 W. This reduction is primarily due to dynamic power scaling down from 0.498 W to 0.197 W, reflecting the inherent sparsity of SNN execution where inactive neurons and untriggered synapses bypass integration steps. Across both architectural versions, static power dissipation remains firmly bounded between 0.138 W and 0.140 W, demonstrating that the structural pipelining applied to the argmax unit in Version 2 improves the maximum frequency without incurring static leakage penalties on the Artix-7 fabric.

\begin{table}[!htbp]
\caption{Detailed On-Chip Power Consumption Breakdown (Vivado Post-Implementation)}
\label{tab:detailed_power_breakdown}
\centering
\resizebox{\columnwidth}{!}{%
\begin{tabular}{lcccc}
\hline
\textbf{Metric} & \begin{tabular}[c]{@{}c@{}}V1 (13.3 MHz)\\Vectorless\end{tabular} & \begin{tabular}[c]{@{}c@{}}V1 (13.3 MHz)\\Vector-based (SAIF)\end{tabular} & \begin{tabular}[c]{@{}c@{}}V2 (167 MHz)\\Vectorless\end{tabular} & \begin{tabular}[c]{@{}c@{}}V2 (167 MHz)\\Vector-based (SAIF)\end{tabular} \\ \hline
\textbf{Total (W)}  & $0.164$ & $0.177$& $0.637$& $0.336$\\
Dynamic (W)         & $0.026$& $0.039$& $0.498$& $0.197$\\
Static (W)          & $0.138$& $0.138$& $0.140$& $0.139$\\
Clocks (W)          & $0.002$& $0.002$& $0.065$& $0.062$\\
Signals (W)         & $0.009$& $0.013$& $0.187$& $0.052$\\
Logic (W)           & $0.014$& $0.023$& $0.241$& $0.081$\\
I/O (W)             & $<0.001$& $<0.001$& $0.005$& $<0.001$\\
\hline
\end{tabular}
}%
\end{table}

\section*{Appendix B: Post-Implementation FPGA Device Floorplans}
\label{sec:appendix_floorplans}

\begin{figure}[!htbp]
    \centering
    \begin{subfigure}[b]{0.51\columnwidth}
        \centering
        \includegraphics[width=\columnwidth]{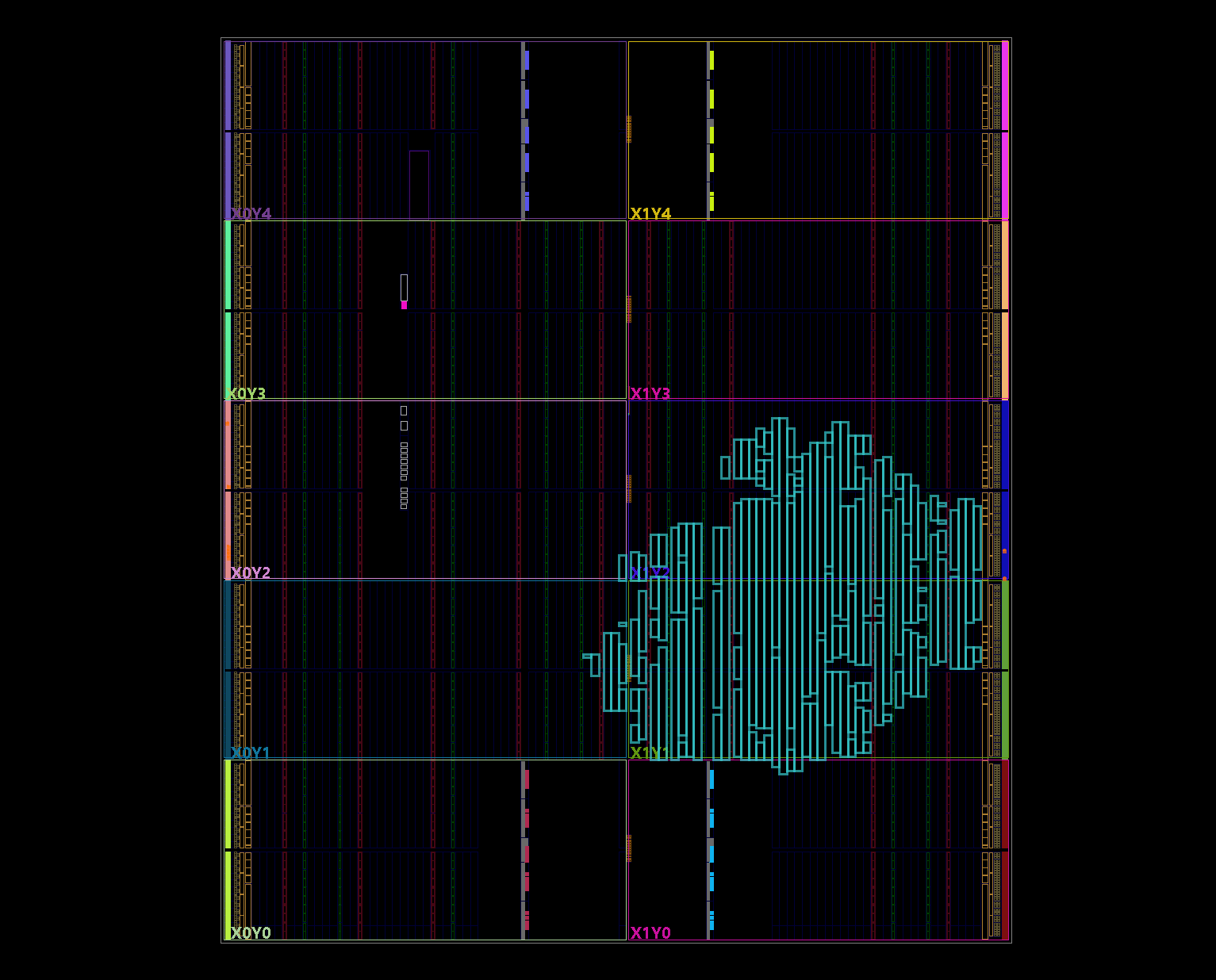}
        \caption{Version 1 Design}
        \label{fig:design_v1}
    \end{subfigure}
    \hfill
    \begin{subfigure}[b]{0.51\columnwidth}
        \centering
        \includegraphics[width=\columnwidth]{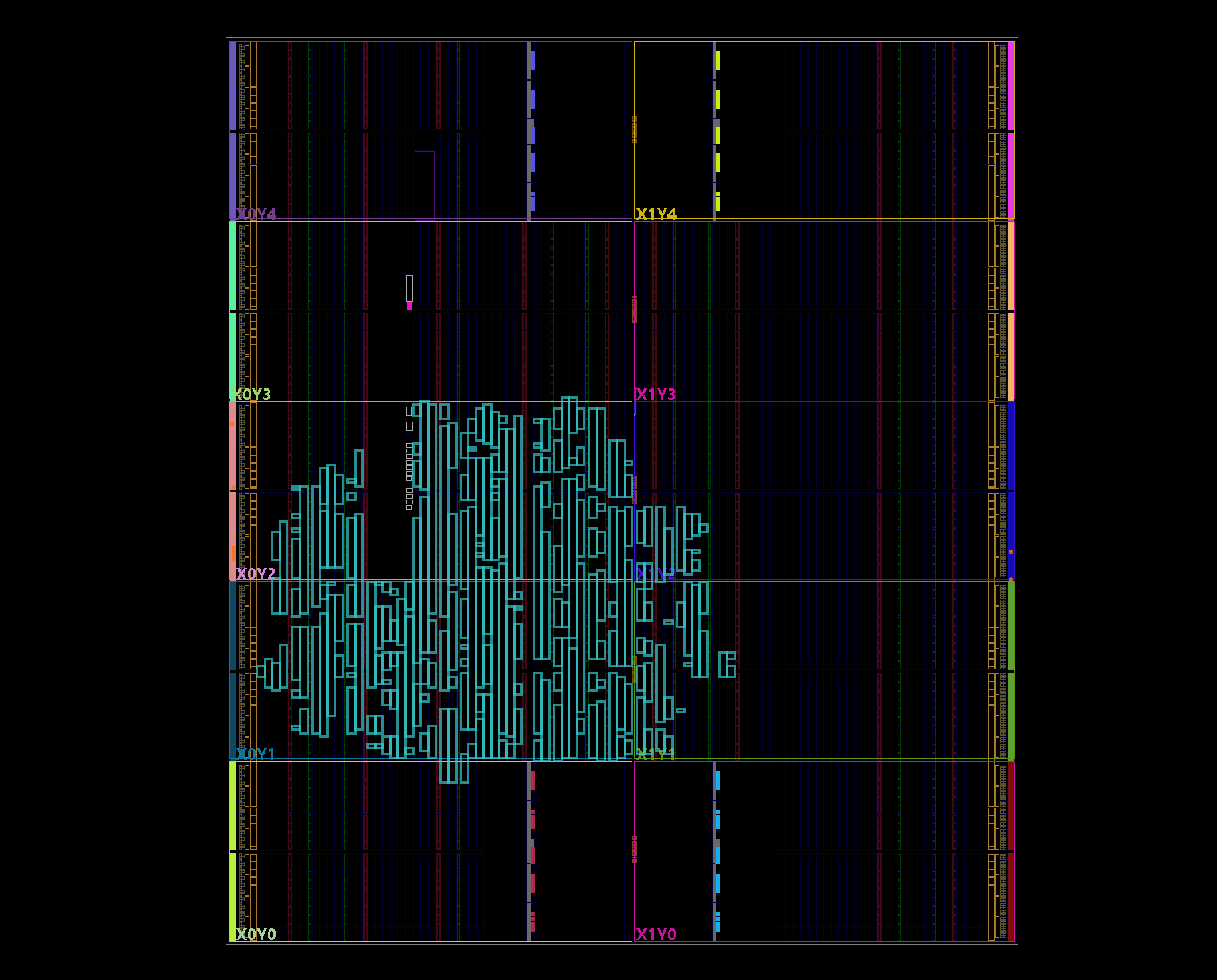}
        \caption{Version 2 Design}
        \label{fig:design_v2}
    \end{subfigure}
    \caption{Physical layout footprints of V1 and V2 architectures.}
    \label{fig:design_comparison}
\end{figure}

\end{document}